\documentclass[11pt,english]{article}

\usepackage[maccyr]{inputenc}
\usepackage{amssymb,amsmath}
\usepackage{amsthm}
\usepackage{cite}
\usepackage{}
\usepackage{bbm}
\usepackage{setspace}
\usepackage{empheq}
\usepackage{fancybox}
\usepackage{tikz}
\usepackage{lscape}
\usetikzlibrary{matrix,arrows,decorations.pathmorphing}

\def\hybrid{
        \topmargin -20pt
        \oddsidemargin 0pt
        \headheight 0pt \headsep 0pt
        \textwidth 6.25in 
        \textheight 9.5in 
        \marginparwidth .875in
        \parskip 5pt plus 1pt \jot = 1.5ex}

\hybrid

\DeclareMathOperator{\sgn}{sgn}

\def\be{\begin{equation}}
\def\ee{\end{equation}}
\def\ba{\begin{align}}
\def\ea{\end{align}}

\begin{document}


\begin{flushright}
CPHT-RR046.072020 \\
DESY 20-123\\
\end{flushright}
\vskip .8 cm
\begin{center}
{\Large {\bf  String Defects, Supersymmetry and the Swampland   }}\\[12pt]

\bigskip
\bigskip 
{
{\bf{Carlo Angelantonj $^\dagger$}\footnote{E-mail: carlo.angelantonj@unito.it}},
{\bf{Quentin Bonnefoy$^\ddagger$}\footnote{E-mail: quentin.bonnefoy@desy.de }},  
{\bf{Cezar Condeescu$^{*}$}\footnote{E-mail: ccezar@theory.nipne.ro}},
{\bf{Emilian Dudas$^{**}$}\footnote{E-mail: emilian.dudas@polytechnique.edu}}
\bigskip}\\[0pt]
\vspace{0.23cm}
{\it $^\dagger$Dipartimento di Fisica, Universita di Torino,INFN Sezione di Torino and Arnold-Regge Center, Via P. Giuria 1, 10125 Torino, Italy \\ \vspace{0.2cm}
$^\ddagger$ Deutsches Elektronen-Synchrotron DESY, 22607 Hamburg, Germany \\ \vspace{0.2cm}
$^{*}$   Department of Theoretical Physics, 
 ``Horia Hulubei" National Institute of Physics and Nuclear Engineering, P.O. Box MG-6, Magurele - Bucharest, 077125, Jud. Ilfov, Romania  \\ \vspace{0.2cm}
$^{**}$ CPHT Ecole Polytechnique, CNRS, IP Paris,  91128 Palaiseau, France}\\[20pt] 
\bigskip
\end{center}

\begin{abstract}
\noindent

\end{abstract}
Recently, Kim, Shiu and Vafa proposed general consistency conditions for six dimensional supergravity theories with minimal supersymmetry coming from couplings to strings.  
We test them in explicit perturbative  orientifold models in order to unravel the microscopic origin of these constraints. 
 Based on the perturbative data, we  conjecture  the existence of null charges $Q \cdot Q=0 $ 
for any six-dimensional theory  with at least one tensor multiplet, coupling to string defects of charge $Q$. We  then include the new constraint to exclude some six-dimensional supersymmetric anomaly-free examples that have currently no string or F-theory realization. We also investigate the constraints from the couplings to string defects in case where supersymmetry is broken in tachyon free vacua, containing non-BPS configurations of brane supersymmetry breaking type, where the breaking is localized on antibranes. In this case, some conditions have naturally to be changed or relaxed whenever the string defects experience supersymmetry breaking, whereas 
the constraints are still valid if they are geometrically separated from the supersymmetry breaking source.    

\newpage

\tableofcontents
\

\section{Introduction and Motivations}

Usually a quantum field theory in an arbitrary dimension is considered to be consistent if at the quantum level all gauge and gravitational anomalies are canceled. However, it is known that most of them cannot
be realized in string theory. More generally, it is believed that some seemingly consistent field theories cannot be coupled to quantum gravity, and belong to the Swampland \cite{swampland}, 
as opposed to the landscape of string theory compactifications. Very often, anomalies are canceled \`a la 
Green-Schwarz \cite{Green:1984sg}: quantum anomalies from chiral fermions or self-dual forms are canceled by tree-level couplings of form potentials of various degrees, which transform nontrivially under
gauge and diffeomorphism transformations. Whereas in perturbative heterotic constructions the Green-Schwarz mechanism is unique in the sense of involving always the universal antisymmetric tensor, 
in type II orientifolds it is non-universal and can involve different tensors of various rank from the Ramond-Ramond (RR) sector, as shown by Sagnotti \cite{Sagnotti:1992qw}. 
Whereas a string theory compactification automatically satisfy this generalized Green-Schwarz-Sagnotti (GSS) anomaly cancellation mechanism, it is natural to inquire the reverse question:
is a field theory which  satisfies the GSS anomaly cancellation conditions always consistent in the UV? The fact that a large class of six dimensional theories compatible with the GSS mechanism could not be realized
explicitly in string or F-theory  \cite{Kumar:2010ru} makes such examples particularly puzzling.  

Recently,  consistency of strings defects  coupled to tensor fields of the theory were worked out \cite{Kim:2019vuc} -\cite{Couzens:2020aat} based on generic properties of the  conformal field theory (see e.g.
 \cite{DiFrancesco:1997nk}) on the string like defects and the anomaly inflow mechanism \cite{inflow},\cite{Douglas:1995bn}, in theories with $(1,0)$ supersymmetry in 10d and 6d  and then generalized to other dimensions\footnote{Some related work include \cite{Kumar:2009ac} -\cite{Kimura}.}.  
The idea is to assume that the spectrum of quantum gravity is complete and use the consistency of theories living on probe branes (in our case string defects) to obtain further swampland constraints in addition to the ones arising from 6d anomaly considerations \cite{Kim:2019vuc}. In the case under study, the conditions 
are essentially on the charge vectors $Q_\alpha$ of string defects,  where $\alpha = 0, \hdots, N_T$, where $N_T$ is the number of tensor multiplets.
If it is not possible to find vectors  $Q$ satisfying a set of constraints, the 6d model belong to the swampland, even if it satisfies all 6d GSS anomaly cancellation conditions.    
If the original goal of \cite{Kim:2019vuc} was to provide conditions for a generic theory of quantum gravity coming from couplings to 2d string defects, it is important to check the generality
of such constraints in known string constructions, with different amount of supersymmetries, especially with little or no supersymmetry.  One of the reasons of doing this is to investigate if there
are assumptions needed in deriving the constraints, if  some of them can be relaxed or if there are other potential constraints needed to be imposed.  We are interested in what follows in testing and generalizing the swampland conditions introduced in \cite{Kim:2019vuc} for 6d theories by explicitly looking at orientifold compactifications of string theory.

On the other hand, it was shown in orientifold compactifications that certain supergravity (closed string) spectra  necessarily break supersymmetry \cite{bsb} at the perturbative level. In the simplest 6d example, the spectrum contains seventeen tensor multiplets, whereas supergravity couples to D-branes with supersymmetry nonlinearly realized on their worldvolume \cite{Dudas:2000nv}.  In such constructions
supersymmetry breaking is localized on a set of (anti)branes,  whereas far from them supersymmetry is still present for the massless excitations. 
The simplest settings of such {\it brane supersymmetry breaking} (BSB) constructions have gauge theory (open string) spectra with fermionic spectrum and anomaly polynomial exactly as a supersymmetric
theory, whereas actually the massless (open string) bosonic partners are in different representations, breaking therefore supersymmetry at the string scale. The necessity of breaking supersymmetry in such models 
is simply due to the fact that an alternative orientifold projection generates exotic $O_+$-planes with positive tension and charge. From this viewpoint, these models are similar to the ten dimensional example
constructed in \cite{sugimoto}. However, in the 10d example the same closed string spectrum can couple to the supersymmetric $SO(32)$ type I superstring or to its non-supersymmetric $USp(32)$ cousin
\cite{sugimoto}.
On the other hand,  the BSB 6d example is a compactification of the type I string, but the corresponding supergravity cannot be coupled to any supersymmetric D-branes, because in  perturbation theory, in 
this case RR tadpole conditions can only be satisfied by adding $\overline{\text{D5}}$ antibranes instead of D5 branes\footnote{Interestingly enough, for our main 6d example,  in was  shown  \cite{Morrison:1996pp}, \cite{Blum:1996hs}  to be possible to have a supersymmetric model in F-theory \cite{Vafa:1996xn} with the same supergravity spectrum, which is an isolated vacuum with no possible deformation parameters.}. The Op$_+$-$\overline{\text{Dp}}$ system is similar to a brane-antibrane configuration from the viewpoint of breaking supersymmetry, 
but crucially it lacks the tachyonic instability of the latter.  A detailed understanding of this models, in particular the ground state and their stability are still not completely settled  \cite{Dudas:2000ff}, 
\cite{Schwarz:2001sf},  \cite{Mourad:2016xbk}. For recent work on the effective field theory with nonlinear supersymmetry and supergravity realization of such models see \cite{Dudas:2015eha}; 
for recent similar string and brane constructions, see  \cite{kallosh}.   
Since in BSB models supersymmetry breaking is geometrically localized, it is natural  to ask if and how the conditions found in \cite{Kim:2019vuc} capture the essence of this phenomenon.
 Having a different perspective on these peculiar string and F-theory vacua was an important motivation for our investigation. 

In this paper we examine Swampland constraints arising from consistently coupling six dimensional supersymmetric and brane supersymmetry breaking (BSB) supergravity-Yang-Mills systems  to the CFT  string
defects via anomaly inflow considerations.   
We use the known data from string orientifold models and their couplings to various objects carrying string defect charges, such as D1 branes and D5 branes wrapping a four-cycle in the internal space with
internal magnetic fields on their worldvolume.   Our investigation leads to a  
microscopic geometric understanding of the structure of the anomaly polynomial and of the charges of string defects.  We find a mini-landscape of string defects of various charges, some of them being non-BPS but stable.
We also define a geometrical factorization of the anomaly polynomial and  show that the integral basis (in the language of  \cite{Kumar:2010ru}) 
for writing the string charges is not the most natural one from the viewpoint of the
geometry of D-branes and their couplings to the tensors, but it is generically related to it by a rotation of charges.  
In all the perturbative construction we considered in this paper and other cases we checked in the literature of 6d models with at least one tensor multiplet,  it was always possible to introduce consistently coupling to specific string defects with charges satisfying $Q \cdot Q =0$, which  we call null charged strings. Whereas we do not have a fully general argument for their existence in all possible 6d vacua, we conjecture their existence as a new consistency test.

On the other hand, one finds that the constraints \cite{Kim:2019vuc} on the string charges are valid in supersymmetric models for the
BPS defects, whereas they are violated on non-BPS (but otherwise stable) charged defects.   Whereas for a generic non-supersymmetric construction it is not clear what type of constraints can be proposed,
one finds that when supersymmetry is locally broken as in BSB constructions, the constraints \cite{Kim:2019vuc} are valid if the string defects are geometrically separated from the source of supersymmetry breaking.
Conversely, if there is no geometric separation and string defects experience supersymmetry breaking in the spectrum, some constraints can be relaxed, in a way compatible with the microscopic perturbative construction.  

Our paper is organized as follows.  
Section 2 contains a short summary of general features of six dimensional models, the anomaly inflow on string defects and proposes a new consistency condition, based on the existence of a special type of string defects, which in a perturbative construction includes mobile (or bulk) branes. We conjecture that such strings, whose charges satisfy $Q \cdot Q =0$, should exist in any consistent 6d theory with at least one tensor multiplet.   
Section 3 describes the geometrical factorization of the anomaly polynomial, which encodes the geometry of D-branes and O-planes.
We then discuss the simplest 6d orbifold supersymmetric example, the Bianchi-Sagnotti-Gimon-Polchinski (BSGP) model \cite{Bianchi:1990yu}, which allows us to
clarify some microscopic interpretations of the consistency constraints in  \cite{Kim:2019vuc}. 
Section 4 introduces the class of models with the phenomenon of brane supersymmetry breaking. We investigate the fate and modification of the constraints in the presence of supersymmetry breaking. 
Both in Sections 3 and 4 we define a large class of models and string defects charges, where the string defects are either D1 branes or D5 branes wrapping a four cycle in the internal space. 

We end with a summary of our results. The Appendices contain our conventions and formulae for the anomaly computations, the string amplitudes that we used to find all spectra of string vacua and
string defects  in the main text  and two extra 6d models related by continuous deformations to our main examples. 

\section{Six Dimensional Supersymmetric Models}

Six-dimensional $\mathcal{N}=(1,0)$ supersymmetry is strongly constrained by anomalies. 
There are four types of supersymmetry multiplets appearing in 6d $\mathcal{N}=(1,0)$ theories: gravity, vector, hyper and tensor multiplets. Their contributions to the anomaly polynomial\footnote{Unless specified, we are only concerned in all our paper with the non-abelian anomalies. The abelian gauge factors and their anomalies do not affect our considerations and results.} 
is summarized in Table \ref{sugra-anomaly}.
\begin{table}[h!]
\centering
\begin{tabular}{cc}
{\bf SUSY Multiplet} & {\bf Anomaly Polynomial}\\
\hline\hline\\[-4pt]
  Gravity  & $-\frac{273}{360} \text{tr} R^4 + \frac{51}{288} \left(\text{tr} R^2 \right)^2$ \\ [8pt]
  Vector & $-N_V\left[\frac{1}{360} \text{tr} R^4 + \frac{1}{288} \left(\text{tr} R^2 \right)^2 \right]-\frac{1}{24} \text{Tr}_{\text{Adj}}F^4+ \frac{1}{24} \text{Tr}_{\text{Adj}}F^2 \, \text{tr} R^2$ \\[8pt]
  Hyper & $N_H\left[\frac{1}{360} \text{tr} R^4 + \frac{1}{288} \left(\text{tr} R^2 \right)^2 \right]+\frac{1}{24} \text{Tr}_{\text{R}}F^4- \frac{1}{24} \text{Tr}_{\text{R}}F^2 \, \text{tr} R^2$\\[8pt]
  Tensor & $\frac{29}{360} \text{tr} R^4  - \frac{7}{288} \left(\text{tr} R^2 \right)^2$\\[8pt] \hline
\end{tabular}
\caption{Contributions to the anomaly polynomial from the various multiplets of 6d $\mathcal{N}=(1,0)$ supergravity. The signs reflect the chirality and duality properties of the field content of the multiplets. $\text{Tr}_{\text{R}}$ refers to a trace in the representation R, and Adj refers to the adjoint representation.}\label{sugra-anomaly}
 \end{table}

It is easy to see from  the same table that the cancellation of the irreducible gravitational anomaly, corresponding to the term $\text{tr} R^4$ in the anomaly polynomial, yields the constraint
 \be
 N_H - N_V = 273-29 N_T \ , \label{gravity}
 \ee
where $N_H$ denotes the number of hypermultiplets, $N_V$ the number of vector multiplets and $N_T$ the number of selfdual or anti-selfdual tensor multiplets.  One can use the identity in eq. \eqref{gravity} 
to  write the generic anomaly polynomial for supersymmetric $\mathcal{N}=(1,0)$ six-dimensional models as follows
 \be
 I_8 = \frac{9-N_T}{8} \left(\text{tr} R^2 \right)^2 - \frac{1}{24} \text{tr} R^2 \text{Tr}_{\psi}F^2 + \frac{1}{24} \text{Tr}_{\psi} F^4  \ , \label{genpoly}
 \ee
where $\text{Tr}_\psi$ denotes a trace over the charged states of a generic model, to which charged hypermultiplets contribute with a plus sign and vector multiplets with a minus sign. Recall that the Green-Schwarz-Sagnotti mechanism in 6d requires a factorization of the form
\be
I_8 = \frac12 \Omega_{\alpha \beta} \, X_4^\alpha \, X_4^\beta \ , \label{anpol}
\ee
where $\Omega_{\alpha \beta}$ can be chosen (by a rotation) to be diagonal of signature $(1,N_T)$.  The polynomials $X_4^\alpha$ are parametrized in terms of $(1+N_T)$-dimensional vectors $a,b_i$ (with $i$ labeling the gauge group factors)\footnote{ From now on, all anomaly polynomials are expressed in terms of the traces $\text{tr}$ in the fundamental representations of the corresponding gauge groups.}
\be
X_4^\alpha = \frac12 a^\alpha \text{tr} R^2 + \frac12 \sum_{i} \frac{b_i^\alpha}{\lambda_i} \text{tr} F_i^2  \ , \label{X4}
\ee
such that we can write
\be
I_8 = \frac18 a\cdot a \left(\text{tr} R^2 \right)^2 + \frac18 \sum_{i,j} \frac{b_i\cdot b_j}{\lambda_i \lambda_j} \text{tr} F_i^2\, \text{tr} F_j^2 + \frac14 \sum_i \frac{a \cdot b_i}{\lambda_i} \text{tr} R^2 \, \text{tr} F_i^2 \ , 
\label{polyab}
\ee 
where the dot products involve the symmetric form $\Omega_{\alpha \beta}$; $a \cdot b_i \equiv a^{\alpha} \Omega_{\alpha \beta} b^{\beta}_i$, etc. The group theory factors $\lambda_i$ in Table \ref{lambda} are chosen such that one obtains integral scalar products $a\cdot a, a\cdot b_i, b_i \cdot b_j \in \mathbb{Z}$ (see \cite{Kumar:2010ru}). The integrality of the lattice generated by $b_i$ can be inferred to be necessary from the Dirac quantization conditions for dyons (see \cite{Deser:1997se},\cite{Seiberg:2011dr}). 

\begin{table}[h!]
\centering
\begin{tabular}{cccc}
{\bf Group} & $SU(N)$ & $SO(N)$ & $USp(N)$\\
\hline \hline\\[-10pt]
$\lambda$ & $1$ & $2$ & $1$\\
\hline
\end{tabular}
\caption{Normalization for the factors $\lambda_i$. } \label{lambda}
\end{table}

The values in Table 2 are guaranteeing in all cases the integrality of the scalar products  $a\cdot a, a\cdot b_i, b_i \cdot b_j \in \mathbb{Z}$, which define an integral lattice.

When the spectrum is such that there is a six-dimensional anomaly, it can be cancelled by adding a tree-level Green-Schwarz term of the form
\be
S_{GS} = \int \Omega_{\alpha \beta}\, C_2^\alpha\wedge X_4^\beta \ , 
\label{GScoupling}
\ee
where $C_2^\alpha$ are the two-forms of the theory which shift under gauge and gravitational transformations. The gauge invariant field strength which appears in the action is modified to include a Chern-Simons three-form (gauge and gravitational) with non-trivial transformation properties  
\be
H_3^\alpha = dC_2^\alpha + \omega_3^\alpha \ , \quad d \omega_3^\alpha \equiv X_4^\alpha \ , \quad \delta_\theta\,  \omega_3^\alpha = \text{tr} \left[d \left(\theta X_2^{1 \alpha} \right) \right] \ ,
\ee
so that the transformations of the two-forms,
\begin{align}
\delta_\theta \, C_2^\alpha = - \text{tr} \left(\theta X_2^{1\alpha} \right) \ , 
\end{align}
induce a classical shift of \eqref{GScoupling} that cancels the anomalous shift of the action due to the spectrum. 

\subsection{String defects and the anomaly inflow}

Demanding a consistent coupling of the theory to  BPS strings, as required by the completeness principle of quantum gravity, gives rise to additional consistency conditions to be imposed \cite{Kim:2019vuc}. 
 Strings (described by D1 branes or D5 branes wrapping a four cycle in the type I string setup of our paper)  coupled to the tensors in six dimensions have charges described by an $N_T+1$ dimensional vector $Q$. 
 Their couplings are given by 
\be
S_{2d} \supset -\, \Omega_{\alpha \beta}\, Q^\alpha \int C_2^\beta \ . 
\label{inflowCoupling}
\ee
The anomalous gauge and gravitational shifts of \eqref{inflowCoupling} add up to the ones of the 2D CFT on the string defect, such that the overall anomaly vanish. It follows then that 
the anomaly polynomial of the two dimensional CFT of the strings is generically of the form
\be
I_4 = \Omega_{\alpha \beta} \, Q^\alpha \left( X_4^\beta  + \frac12 Q^\beta \chi(N) \right) \ , \label{polyQ}
\ee
where $\chi(N)$ is the Euler class of the normal bundle of the string (with worldvolume embedded into 6d spacetime). One can further write the polynomial above in terms of the products $Q \cdot a$, $Q \cdot b_i$ and $Q \cdot Q$ by making use of the general form of $X_4^\alpha$ in eq. \eqref{X4}
\be
I_4 = \frac12 Q \cdot a\, \text{tr} R^2 + \frac12 \sum_i \frac{Q \cdot b_i}{ \lambda_i} \text{tr} F_i^2 + \frac12 Q \cdot Q\, \chi(N) \ . \label{polyQ2}
\ee
The explicit form of the constraints formulated in \cite{Kim:2019vuc}, for a 6d gauge theory of gauge group 
$G = \prod_i G_i$ coupled to (super)gravity, is:
\begin{eqnarray}
Q \cdot J \geq 0 \quad , &&\quad Q \cdot Q \geq -1 \quad , \quad Q \cdot Q + Q \cdot a \geq -2 \ , \quad k_i \equiv Q \cdot b_i \geq 0 \ , \nonumber \\
&&\sum_i \frac{k_i \dim G_i}{k_i + h_i^{\vee}}  \leq c_L  \ , \label{constraints}
\end{eqnarray}
where $k_i$ are the levels of the $G_i$  current algebra, $h_i^{\vee}$ the dual Coxeter number of the gauge group factor $G_i$ and $c_L$ is the central charge for the left-moving sector on the string (D1 branes in
the examples of this section).
For the case of a non-degenerate 2d SCFT on a string,   \cite{Kim:2019vuc} wrote down the explicit expressions for the left/right central charges
\begin{equation}
c_L = 3 Q \cdot Q - 9 Q \cdot a + 2 \quad ,  \quad c_R = 3 Q \cdot Q - 3 Q \cdot a  \ , \label{constraints2}
\end{equation}
whose positivity impose some of the conditions in (\ref{constraints}). The first line of \eqref{constraints} are the conditions that define a would-be consistent string, while the second line is a unitarity constraint that any would-be consistent string must respect according to the completeness principle, otherwise the theory is in the swampland. The K\"ahler form $J$ in  (\ref{constraints}), which is also a $N_T+1$ vector, 
 is constrained by supersymmetry.  Indeed, consistency of the moduli space of scalars lead to the conditions
\begin{align}
J \cdot J &> 0 \ , & J\cdot a &<0 \ , & J\cdot b_i &>0 \ . 
\label{conditionsJ}
\end{align}

\subsection{The Null Charged Strings Conjecture }
\label{sec:null}

In all examples we discuss in the following sections {\it with at least one tensor multiplet} in orbifold orientifold compactifications,  for the bulk D1 branes (and also at fixed points, if they have no twisted charges), 
their charge vector is null $Q\cdot Q = 0$. The microscopic explanation is that in this case D1 branes couple only to the untwisted tensor fields, with equal couplings.  More generally, using the general form
of  (\ref{polyQ}), one can see that the null D1 branes correspond to a world-volume anomaly polynomial $I_4$ which does not depend on the normal bundle.
An obvious particular case is when the worldvolume fermions are non-chiral respect to the $SU(2)_l \times SU(2)_R$ normal bundle. This is for example the case where
the 6d model corresponds to a geometric compactification of a 10d string. For non-geometric compactifications, we are not aware of a simple argument. However, in all the examples
we checked the null charged strings always exist and satisfy all the required consistency conditions\footnote{ In addition to the examples discussed in this paper, we checked the examples in   \cite{Angelantonj:1996mw},
\cite{Bianchi:1999uq},\cite{Blumenhagen:2000fp}. For more standard geometric compactifications like  \cite{Gimon:1996ay}, \cite{Dabholkar:1996pc} their existence is guaranteed due to the  geometric
nature of the compactifications of the 10d type I superstring.}. 
We  conjecture that they should exist in all consistent 6d theories. In what follows 
we check, by using this conjecture, which examples discussed in the literature, compatible with all the other constraints,  could be ruled out by the non-existence of the null charged strings.

The first example we consider is $N_T=1$ with gauge group $SU(N)$ coupled to one symmetric and $N-8$ fundamental hypermultiplets introduced in  \cite{Kumar:2010ru}, \cite{Kumar:2009ac}, \cite{Kim:2019vuc}. This model has no known string or F-theory embedding, and the strongest known constraint is $N \leq 30$ from anomaly cancellation conditions, whereas
the conditions eqs. (\ref{constraints}) are satisfied for $N \leq 117$ \cite{Kim:2019vuc}. The anomaly vectors verify
\be
a\cdot a = 8 \ , \quad a\cdot b=1 \ , \quad b\cdot b = -1 \ .
\ee
When $N_T=1$, there are two possible lattices \cite{Seiberg:2011dr},
\be
\Omega_0=\left(\begin{matrix}0&1\\1&0\end{matrix}\right)\ , \quad \Omega_1=\left(\begin{matrix}1&0\\0&-1\end{matrix}\right) \ ,
\ee
but using $\Omega_0$, one immediately sees that there cannot be any $b=(b_0,b_1)$ with integer entries, since $b\cdot b=2b_0 b_1=-1$. Thus we can concentrate on $\Omega_1$. The only integer-valued $a$ (up to parity transformations in $SO(1,1)$) is $a = (-3,1)$, for which the only integer-valued solution for $b$ is\footnote{For such values, an example of Kahler form is $J=(n,1)$, with $n>0$, but we do not need to discuss the value of $J$ in what follows.} $b = (0,-1)$. If we now consider a null charged string $Q=(\epsilon q, q)$, with $\epsilon=\pm 1$, and impose the subset of constraints (\ref{constraints}) $k,k_l,c_R\geq 0$, we find that $q\geq 0$ and $0\leq (3\epsilon +1)q\leq 2$, which cannot be satisfied for a non-zero integer $q$. Hence, there does not exist any consistent charge vector for a null string and therefore the model can be excluded by the inconsistency of the coupling to null charged strings. 

Our second example from \cite{Kumar:2010ru} has $N_T=1$, gauge group $G = SU(24) \times SO(8)$ and three matter hypermultiplets in the antisymmetric representation of the unitary gauge factor. 
This model has also no known F-theory embedding and the reason for this was explained in \cite{Kumar:2010ru}. From a more general supergravity viewpoint however, it satisfies all the consistency conditions
eqs. (\ref{constraints}).  
In  this model, the anomaly vectors verify\footnote{   Our $b_2$ seems to have twice the entries compared to \cite{Kumar:2010ru}. This is probably due to a different convention for the traces in the
$SO$ gauge group factors. We remind that ours is that all anomaly polynomials are always expressed in terms of the traces in the fundamental representations of the corresponding gauge groups.}
\be
a\cdot a = 8 \ , \quad a\cdot b_1=-3 \ , \quad a\cdot b_2=2 \ , \quad b_1\cdot b_1 = 1 \ ,\quad b_2\cdot b_2 = 4 \ ,  \quad b_1\cdot b_2 = 0 \ ,
\ee
and again $\Omega_0$ cannot lead to any solution with integer $b_{1,2}$. Using $\Omega_1$, the most general solution (up to parity transformations in $SO(1,1)$ again) is\footnote{The Kahler form could be chosen to be $J = (2,1)$.}
\be
\quad a = (-3,1) \ ,   \quad b_1 = (1,0) \ ,  \quad b_2 = (0,-2) \ . \label{null2}
\ee
Similar to the previous example, using $k_1,k_2,k_l,c_R\geq 0$, null strings $Q=(\epsilon q,q)$ in this case have to verify $q\geq 0,\epsilon=1$ and $0\leq (3\epsilon +1)q\leq 2$. There are therefore no consistent charge vectors
for null strings coupling to tensors and the model can be excluded by the null charged string hypothesis.

Since we do not have a fully general argument, the null strings hypothesis could be violated in some exotic 6d theories. If such examples would be found in string theory, they would probably correspond to truly
non-geometric compactifications.   

\section{Six Dimensional Supersymmetric Orientifold Models}

Whereas the main point in \cite{Kim:2019vuc} was  trying to  use only general arguments, valid beyond string perturbation theory, the opposite viewpoint, analyzing explicit perturbative examples has its own virtues. 
Indeed, this can offer a microscopic insight on such constraints, delimitate their generality, for example by relaxing supersymmetry and exploring known examples.  It can also help to identify and test
new constraints which are  suggested by the perturbative string data, in particular our conjecture on the existence of the null charged strings $Q \cdot Q =0$.  We consider therefore supersymmetric orbifold 6d examples
in what follows. Before turning to explicit examples, we introduce and discuss a natural factorization of the anomaly polynomial in six dimensional perturbative models, called geometric in what follows, determined by the geometry of the branes and O-planes and their couplings. 

\subsection{Geometric factorization of the anomaly polynomial }

The anomaly polynomial  (\ref{anpol}) is a $(N_T+1) \times (N_T+1) $ quadratic form in $n_G+1$ variables, where $n_G$ is the number of non-abelian gauge group factors, if one ignores, as we already stressed,
the abelian factors. If $N_T \leq n_G$, the factorization of the polynomial is uniquely determined by the diagonalization of the quadratic form when the eigenspaces are of dimension one. If  $N_T > n_G$, the factorization cannot be unique and one should use further insight in order to find the most relevant
factorizations.  The mapping between various factorizations is obtained by performing $SO(1,N_T)$ transformations, which amount to redefining accordingly the basis of tensor fields participating in the anomaly cancellation mechanism. One particularly important factorization, that we call geometric in what follows, is the one reflecting the geometry of the D-branes and O-planes in the internal space and their couplings to the 
(closed string) tensors, in perturbative constructions. This could be called the string basis, since the associated tensors basis is the one corresponding to perturbation theory, whereas all other basis are obtained by taking linear
combinations of these tensors. 

Due to the interpretation of the Klein, cylinder and Mobius string amplitudes as tree-level closed string exchanges between D-branes and O-planes, the geometric basis can be found most easily from the 
vacuum-to-vacuum string partition functions, restricted to the massless tree-level RR (tadpole) exchange. The corresponding amplitudes for six-dimensional constructions have typically the structure
\begin{align}
&- {\tilde {\cal K}}_0 - {\tilde {\cal A}}_0 - {\tilde {\cal M}}_0 = \left[  \left(\sum_i \alpha_i N_i - 32\right) \sqrt{v} + \epsilon  \frac{\sum_a \alpha_a D_a - 32 }{\sqrt{v}}  \right]^2 \chi_{+} \ \nonumber \\
& +  \left[  \left (\sum_i { \alpha}_i N_i - 32\right ) \sqrt{v} - \epsilon  \frac{\sum_a \alpha_a D_a - 32 }{\sqrt{v}}  \right]^2 \chi_-   + \sum_{\alpha'} \left(  \sum_i c_i^{\alpha'} N_i+   \sum_a c_a^{\alpha'} D_a 
-   c_O^{\alpha'} \right)^2  \chi^{(\alpha')}  \ . \label{gf1}
\end{align}
In  (\ref{gf1})  $(\chi_+,\chi_-, \chi^{(\alpha')}) $ are a basis of $N_T+1$ group characters corresponding to RR six-forms, enforcing the RR tadpole cancellations.   $\chi_+,\chi_-$ are  characters that, at the massless level, contain untwisted six-forms, whereas  in orbifold models $\chi^{(\alpha')}$ correspond to twisted RR six-forms. 
If the gauge group is of the form $G = \prod_i G_i \otimes \prod_a G_a$, $N_i$ is proportional to the number of branes of the D9 gauge group factor $G_i$ and $D_a$ to the number of branes of the D5 gauge group factor $G_a$. Moreover, 
$v$ is the volume of the internal space, $\epsilon = \pm 1$ if D5/O5  branes/planes are present and $\epsilon=0$ otherwise, $c_i^{\alpha'}$ ($c_a^{\alpha'}$) denotes couplings of D9 (D5) branes to the twisted sector,
whereas $c_O^{\alpha'}$ is the twisted charge of the O-planes coincident with orbifold fixed points.  In the simplest cases $\alpha_i  = {\alpha}_a=1$ corresponding to rank 16 gauge factors for D9
 and D5 branes  (if present), but  gauge groups of lower ranks can be obtained in various ways.  

The amplitudes above contain the square of the couplings of the supergravity fields and the RR forms to D-branes \cite{Douglas:1995bn} and O-planes \cite{Morales:1998ux}.  From the effective field theory viewpoint, these couplings are encoded into compact formulae containing also the couplings of lower-dimensional tensors to gauge fields and to the gravitational sector. For D9 branes and the O9 plane, the couplings to the untwisted RR fields (those of the 10d gravity multiplet) are captured by   
\begin{align}
 - S_{D9} &= {\text tr} \int_{R^{1,5} \times T^4} C \wedge e^{i F_9} \wedge \sqrt{{\hat A}(R)} \  , & - S_{O9} &= -32  \ {\text tr} \int_{R^{1,5} \times T^4} C \wedge \sqrt{{L}\left(\frac{R}{4}\right)}   \ ,  \label{gf2}
\end{align}
where the formulae for the roof genus ${\hat A}(R)$ and Hirzebruch polynomial $L$ are given in Appendix \ref{appA} and $C$ denotes the formal sum of all (untwisted) RR-fields, $C\equiv C^{(10)}+C^{(6)}+ C^{(2)}$. 
We also assume a compactification on a flat four-dimensional torus $T^4$ (or orbifolds, as in our string examples).  The compactified theory lives in the non-compact 6d Minkowski space $R^{1,5}$, which is the case of interest in this paper.
Similarly, the couplings to the untwisted  RR fields for D5 branes and (one) O5 plane are captured by
\begin{align}
 - S_{D5} &= {\text tr} \int_{R^{1,5}} C \wedge e^{i F_5} \wedge \sqrt{{\hat A}(R)}   \ , &  - S_{O5} & = -2 \ {\text tr} \int_{R^{1,5}} C \wedge \sqrt{{L}\left(\frac{R}{4}\right)}   \ .  \label{gf3}
\end{align}
One finds
\begin{align}
  - (S_{D9}+S_{O9}) &=  \int_{R^{1,5} \times T^4} \left\{ (N-32) C^{(10)} + C^{(6)} \wedge \left(  \frac{N+16}{24}  {\text tr} R^2 - \frac{1}{2}  {\text tr} F_9^2  \right) + \cdots   \right\} \ ,  \nonumber \\
 & =    \int_{R^{1,5}} \left\{ (N-32)\,  v\,  {\tilde C}^{(6)} + {\tilde C}^{(2)} \wedge \left(  \frac{N+16}{24}  {\text tr} R^2 - \frac{1}{2}  {\text tr} F_9^2  \right) + \cdots \right\} \ , 
  \label{gf4}
\end{align}
$N$ is twice the number of D9 branes\footnote{Our terminology here is that the number of branes is equal to the rank of the gauge group, therefore sixteen. Actually in our particular SUSY example, if the D9 branes have 
continuous Wilson lines in $T^4$, their number is divided by two and equals eight, plus their images. } and where $v\,  {\tilde C}^{(6)} \equiv \int_{T^4} C^{(10)}$, ${\tilde C}^{(2)} \equiv \int_{T^4} C^{(6)}$. Similarly, 
\be
 - (S_{D5}+16 S_{O5}) =    \int_{R^{1,5}} \left\{ (D-32) {C}^{(6)} + {C}^{(2)} \wedge \left(  \frac{D+16}{24}  {\text tr} R^2 - \frac{1}{2}  {\text tr} F_5^2  \right)  \right\} \ , 
  \label{gf5}
\ee
where $D$ is twice the number of D5 branes\footnote{Like for D9 branes, the number of D5 branes is equal to the rank of the gauge group, therefore sixteen. If the D5 branes are  off the orbifold fixed points, their number is divided by two and equal eight, plus their orbifold images. }. The  connection with the tadpole amplitudes  (\ref{gf1}) is made transparent by defining linear combinations 
\begin{align}
 \sqrt{v} \ C^{(6)} &= C^{(6)}_+ -  C^{(6)}_-\ , &  \sqrt{v} \ {\tilde C}^{(6)} &= C^{(6)}_+ +  C^{(6)}_- \ , \nonumber \\
 C^{(2)} &= C^{(2)}_+ +  C^{(2)}_- \ , &  {\tilde C}^{(2)} &= C^{(2)}_+ -  C^{(2)}_- \  , \label{gf6}
\end{align}
where $C^{(2)}_+$  ($C^{(2)}_-$) correspond to the self-dual (anti self-dual) untwisted tensor in 6d, whereas $C^{(6)}_+$, $C^{(6)}_-$ are more appropriately interpreted as the string basis for the non-propagating six forms, enforcing the untwisted RR tadpole conditions. Notice that the gauge traces above are written in the $SO(32)$ basis. 

Our discussion until now was general. In order to be more explicit, we consider the 6d SUSY example discussed in the next section, which has sixteen D9 and D5 branes and a gauge group  $G = U(16)_9 \times U(16)_5$.  To express the traces above in the $U(16)$ basis, one can use, for both D9 and D5 branes,
the equalities $ ({\text tr} F^2)_{SO(32)} = 2 ({\text tr} F^2)_{U(16)} $. In the $U(16)$ basis, one finally obtains 
\begin{align}
 & - S_{D9+O9+D5+16 O5} =   \int_{R^{1,5}} \left\{ \left[ \sqrt{v} (N-32) + \frac{D-32}{\sqrt{v}}  \right]  {C}_+^{(6)} +   \left[ \sqrt{v} (N-32) - \frac{D-32}{\sqrt{v}}  \right]  {C}_-^{(6)}
 \right.   \nonumber \\
 & \left.   + C_+^{(2)} \wedge \left(  \frac{N+D+32}{24}  {\text tr} R^2 - \frac12{\text tr} F_9^2 -  \frac12{\text tr} F_5^2 \right)    +   C_-^{(2)}\wedge \left( \frac{D-N}{24}  {\text tr} R^2 +  \frac12{\text tr} F_9^2 -  \frac12{\text tr} F_5^2 \right)  
 +   \cdots \right\} \ . 
  \label{gf7}
\end{align}
The couplings of the two-forms cancel the 6d anomaly polynomial, as discussed in \eqref{GScoupling}, and they are given by the expressions above. It is then clear that the tadpole conditions, encoded in the couplings of the six-forms and which impose $N=D=32$, fix the geometric structure of the couplings of D-branes and O-planes to the tensor fields that determine the anomaly polynomial.
  
There are also couplings of D-branes and O-planes to the other tensors from the twisted sectors, the last term in  eq. (\ref{gf1}). They are described by couplings qualitatively similar to   (\ref{gf2})-(\ref{gf3}), involving the action of the orbifold on the Chan-Paton factors. The logic leading to the connection between the (twisted, in this case) tadpole conditions and the couplings to tensors in the anomaly polynomial is the same as for the untwisted sector. 
Since some details are model-dependent, we do not attempt here to write explicitly these couplings. 


\subsection{An explicit example : the $T^4/\mathbb{Z}_2$ orientifold model}

Arguably the simplest and most popular chiral $\mathcal{N}=(1,0)$ supergravity can be obtained in perturbative string theory by compactifying Type I theory on a $T^4/\mathbb{Z}_2$ orbifold with standard O9/O5 planes (that is negative tension and charge) \cite{Bianchi:1990yu}, constructed first by
Bianchi and Sagnotti and interpreted geometrically later by Gimon and Polchinski.  The closed string part of the spectrum is given in Table \ref{Osusy}.  Four of the hypers come from the untwisted sector, whereas the other sixteen come from the twisted sector, one per fixed point. 
\begin{table}[h!]
\centering
\begin{tabular}{ccc}
{\bf Multiplicity} & {\bf Multiplet} & {\bf Field Content}\\
\hline \hline\\[-10pt]
1& Gravity & $ (g_{\mu \nu}, C_{\mu\nu}^+,\psi_{\mu L})$ \\[2pt]
1 & Tensor & $(C_{\mu\nu}^-,\phi,\chi_R)$ \\[2pt]
 20 & Hypers &  $(4 \phi_a, \psi_{aR})$\\[2pt]
 \hline
\end{tabular}
\caption{Closed string spectrum for the $T^4/\mathbb{Z}_2$ orientifold with O9$_-$/O5$_-$ planes. We have indicated the (on-shell) field content of each multiplet together with the chirality $L,R$ for the fermions and duality $\pm$ for the tensor fields.} \label{Osusy}
\end{table}

In order to cancel the tadpoles generated by the O9/O5 planes, one needs to introduce an open string sector. The simplest solution contains two stacks:  16 D9 branes and 16 D5 branes sitting at a given fixed point of the orbifold ({\it e.g.} the origin of the lattice), such that the gauge group is
\be
G = U(16)_9 \times U(16)_5 \ . 
\ee
The open string spectrum of this solution consists of vector multiplets and hypermultiplets in the representations indicated in Table \ref{openS}.
\begin{table}[h!]
\centering
\begin{tabular}{ccc}
{\bf Multiplet} & {\bf Field Content} & {\bf Representation}\\
\hline \hline\\[-10pt]
Vector &  $(A_\mu, \chi_L)$ & $(256,1)+(1,256)$\\[2pt]
Hyper & $ (4\phi, \chi_R)$  & $(120+\overline{120},1) + (1,120+\overline{120}) +(16,16)$\\[2pt]
\hline
\end{tabular}
\caption{Open string spectrum for the $T^4/\mathbb{Z}_2$ orientifold with O9$_-$/O5$_-$ planes.} \label{openS}
\end{table}

One can show that the anomaly polynomial for the $U(16)_9 \times U(16)_5$ model has the following factorized expression
\be
I_8 = \frac{1}{16} \left(-4\, \text{tr} R^2 + \text{tr} F_1^2 + \text{tr} F_2^2 \right)^2 - \frac{1}{16} \left(-\text{tr} F_1^2 + \text{tr} F_2^2 \right)^2 \ , 
\label{factorization}
\ee
where one can read off $X_4^\alpha$ in \eqref{anpol} with $\Omega = \text{diag}(1,-1)$. We call this form of the anomaly polynomial geometrical, since it is what comes naturally from the coupling of the tensor multiplet to the D9 and D5 branes. Indeed, the first term
in  (\ref{factorization}) reflects the coupling of the anti-selfdual tensor from the gravity multiplet to the D9 and D5 branes, whereas the second term is the coupling of the self-dual counterpart.  In this model, there 
are no physical couplings to the twisted sector fields, in other words the branes have no twisted charges and are therefore regular. 
All this is manifest in the string vacuum amplitudes (see \cite{Angelantonj:2002ct})
\begin{align}
&- {\tilde {\cal K}}_0 - {\tilde {\cal A}}_0 - {\tilde {\cal M}}_0 \sim \left[  (n+ {\bar n} - 32) \sqrt{v} +  \frac{ d+ {\bar d} - 32 }{\sqrt{v}}  \right]^2 C_4 C_4\ \nonumber \\
& +  \left[  (n+ {\bar n} - 32) \sqrt{v} -   \frac{ d+ {\bar d} - 32 }{\sqrt{v}}  \right]^2 S_4 S_4   - \left[  (n-{\bar n}+ 4 (d-{\bar d}))^2 + 15  (n-{\bar n})^2 \right]  S_4 O_4 \ ,  \label{fac1}
\end{align}
where the gauge group is parametrized here by $U(n)_9 \times U(d)_5$ and $N = 2n=32$ and $D=2d=32$ by the tadpole conditions.  In (\ref{fac1}) $C_4C_4,S_4 S_4$ are the characters corresponding
to the untwisted six-forms $\chi_+,\chi_-$ in (\ref{gf1}), whereas  $S_4 O_4$  correspond to the twisted six-forms $\chi^{(\alpha')} $.  Since in this case $N_T=1$ and  $n_G=2$, the factorization of the anomaly 
polynomial is completely determined by the diagonalization of the corresponding quadratic form.  

From  (\ref{factorization}) one can read off the vectors $a$ and $b_i$ of \eqref{X4}
\begin{align}
 a&= (-2\sqrt{2}, 0) \ , & b_1 &= \left(\frac{1}{\sqrt{2}},-\frac{1}{\sqrt{2}} \right) \ , &   b_2 & =   \left(\frac{1}{\sqrt{2}},\frac{1}{\sqrt{2}} \right)\ , \label{abdiag}
\end{align}
and find the products
\begin{align}
a\cdot a &= 8 \ , & a\cdot b_1 &= a \cdot b_2 = -2 \ , & b_i\cdot b_j & = \left(\begin{array}{cc}
0 & 1\\
1 &0
\end{array} \right) \ . \label{prodsusy}
\end{align}
The expressions for $a$ and $b_i$ are unique up to overall signs once one chooses the geometrical basis, with $\Omega$ in the diagonal form. This is true because the number of non-zero eigenvalues of  $I_8$, seen as a quadratic from in the variables $\text{tr} R^2, \text{tr} F_1^2, \text{tr} F_2^2$, is equal to $1+N_T = 2$. We will see later when we analyse the non-supersymmetric $T^4/ \mathbb{Z}_2$ orientifold that $N_T+1$ can be larger than the dimension of the eigenspace of $I_8$, thus several factorizations of the anomaly polynomial can arise even after fixing $\Omega$. However, 
the geometry of the branes provides always a natural, geometrical choice for the factorization. Notice that in the geometrical basis above  the vectors $a,b_i$ do not have integral entries. Since the scalar products in (\ref{prodsusy}) are integers, they generate an integral lattice. It has been argued in \cite{Seiberg:2011dr} that one has to be able to embed the integral lattice generated by $a$ and $b_i$ into a selfdual lattice. In order to check this property it is sufficient to find a basis with integer entries. Indeed, this is  realized in our example by going to a basis where $\Omega$ has an off-diagonal form
\begin{align}
\Omega = \left(\begin{array}{cc}
0& 1\\
1 &0
\end{array} \right) \ . 
\label{integralOmegaGP}
\end{align}
In the integral basis the anomaly polynomial is factorized as
\be
I_8 = \left(\text{tr} R^2 - \frac12 \text{tr} F_1^2 \right)  \left(\text{tr} R^2 - \frac12 \text{tr} F_2^2 \right)
\ee  
and the vectors $a$, $b_i$ have the representation
\begin{align}
 b_1 &= \left(1,0 \right)\ , & a&= (-2, -2) \ , & b_2 & =   \left(0,1 \right)\ ,
\label{integralVectorsGP}
\end{align}
with the products given before in eq. \eqref{prodsusy}. 
The two representations of $\Omega$ and $a,b_1,b_2$ are simply related by a rotation matrix
\be
\mathcal{R}=\frac{1}{\sqrt{2}}\left(\begin{array}{cc}
1 & -1\\
1 & 1
\end{array}\right) \ .
\ee
From the explicit form of $a,b_1,b_2$ in eq. \eqref{abdiag} one can immediately see that the Kahler form $J$ must satisfy the conditions
\begin{align}
J_0 &> 0 \ , & |J_0| > |J_1|
\end{align}
in the geometrical basis (with diagonal $\Omega$). A consistent choice is then easily found to be $J= ({\sqrt 2},0)$, which in the integral basis become $J = (1,1)$.

\subsection{D1 branes and the anomaly inflow}

Before studying the inflow for D1 branes in the supersymmetric $T^4/\mathbb{Z}_2$ orientifold considered earlier, let us analyze the consistency of this model following the general prescription of \cite{Kim:2019vuc}. 
Let us choose the integral basis \eqref{integralOmegaGP}-\eqref{integralVectorsGP} (and $J=(1,1)$), and consider a generic string which couples to the tensor fields with charge $Q=(q_0,q_1)\in\mathbb{Z}^2$. 
For our specific case, in the integral basis, $k_1=q_1$, $k_2=q_0$, $Q\cdot Q=2q_0q_1$, $Q\cdot a=-2(q_0+q_1)$, so that all conditions in the first line of \eqref{constraints} are satisfied if $q_0\geq 0$, $q_1\geq 0$ and $\sgn(q_0-1)=\sgn(q_1-1)$. Notice in particular that these conditions force $Q\cdot Q$ and $Q \cdot a$ to be even integers. 
Now, 
\begin{eqnarray}
&& c_L-\sum_i\frac{k_i\, \text{dim}\,G_i}{k_i+h^\vee_i} = \nonumber \\
&& = \frac{2 \left(3 q_0^2 q_1^2+57 q_0^2 q_1+144 q_0^2+57 q_0 q_1^2+802 q_0 q_1+280 q_0+144 q_1^2+280 q_1+256\right)}{(q_0+16)(q_1+16)} 
\end{eqnarray}
which is obviously positive for positive charges. Taking into account the two $U(1)$ factors of the gauge group, there could be at most an extra $-2$ added in the left hand-side \cite{Lee:2019skh}, which does not change the conclusion.
Therefore, as expected, the $T^4/\mathbb{Z}_2$ orientifold can be consistently coupled to charged strings.

We are now looking at the spectrum of D1 branes in the same supersymmetric $T^4/\mathbb{Z}_2$ orientifold with D9/D5 branes. For the sake of generality we look at sectors with gauge group
\be
U(r)_1 \times U(n)_9 \times U(d)_5
\ee
and leave for the time being the numbers of branes $r,n,d$ arbitrary. The corresponding vacuum amplitudes can be found in Appendix \ref{appB}. The spectrum of strings charged under the D1 gauge group is reproduced in Table \ref{susy-fix}. The R-symmetry of the D1 brane CFT is identified with the normal bundle $SU(2)_l \times SU(2)_R$, whereas  $SO(1,1)$ is the Lorentz group on the worldvolume of the D1 brane. Finally, the $SO(4)$ corresponds to the toroidal orbifold directions.
\begin{table}[h!]
\centering
\begin{tabular}{c c}
{\bf Representation} & {\bf $SO(1,1)\times SU(2)_l \times SU(2)_R \times SO(4)$}\\
\hline\hline\\[-10pt]
 $r \bar r$ & $(0,1,1,1)+(\frac12,1,2,2')_L$ \\[2pt]
$r \bar r$ & $(1,2,2,1)+(\frac12,2,1,2')_R$\\[2pt]
$\frac{r(r+1)}{2}+ \frac{\bar r(\bar r +1)}{2}$ & $(1,1,1,4)+(\frac12,1,2,2)_R$\\[2pt]
$\frac{r(r-1)}{2}+ \frac{\bar r(\bar r -1)}{2}$ &$(\frac12,2,1,2)_L$\\[2pt]
$r \bar n+ \bar r n$ & $(\frac12,1,1,1)_L$\\[2pt]
$rd + \bar r \bar d$ & $(\frac12,1,1,2)_L$\\[2pt]
$r \bar d + \bar r d$ & $(1,1,2,1)+(\frac12,1,1,2')_R$\\[2pt] \hline
\end{tabular}
\caption{The spectrum of D1 branes at a fixed point on the $T^4/\mathbb{Z}_2$ supersymmetric orientifold.} \label{susy-fix}
\end{table}

The anomaly polynomial on the D1 brane worldvolume receives the following contributions (see e.g. \cite{Dudas:2001wd})
\be
I_4 =\frac12 \hat A(R) \hat A(N)^{-1} \times \left\{
\begin{split}
&\text{ch}_\pm(N) \, \text{Tr}_\psi e^{iG}\\
&\text{tr}\,  e^{iF}\, \text{tr}\,  e^{iG}
\end{split} \right. \ , \label{polyD1}
\ee
with the factor $1/2$ arising from the fact that the relevant fermions $1/2_L$ and $1/2_R$ of $SO(1,1)$ are Majorana-Weyl. The relevant expansions of the Dirac genus $\hat A$ and the Chern characters $\text{ch}_\pm$ are given in Appendix \ref{appA}. 
From the spectrum above one can see that the anomaly polynomial corresponding to the $r$ D1 branes becomes
\be
I_4 =- r \left(\text{tr} R^2 - \frac12 \text{tr} F_1^2 \right) \ . 
\ee
One matches the polynomial above with the form given in eq. \eqref{polyQ2}. It immediately follows, after setting $r=1$, i.e. considering a single D1 brane, that we must have the following products
\begin{align}
Q\cdot Q &= 0\ , & Q\cdot a &= -2\ , & Q\cdot b_1 & = 1\ , & Q \cdot b_2 &= 0 \ .  
\end{align}
After solving the constraints above, with $a$,$b_i$ given in eq. \eqref{abdiag}, one finds that the tensor charges $Q$ of the D1 branes in the geometrical basis are then given by
\be
Q = \frac{1}{\sqrt{2}} \, (1,1) \label{chargeQ} \ . 
\ee
Notice that we have $Q=b_2$, corresponding to the fact that the D1 brane is equivalent, in the field theory limit, to an instanton for the gauge theory living on the D5 brane worldvolume. In the integral basis (where $\Omega$ is off-diagonal) the charges have the form $Q = \,  (0,1)$. Recall that the Kac-Moody levels of the gauge factors and of the $SU(2)_l$ have  the form 
\begin{align}
k_i &= Q\cdot b_i\ , & k_l & = \frac12 \left(Q\cdot Q + Q\cdot a +2 \right) \ .
\end{align}
 It is now immediate to obtain the following relevant quantities 
\begin{align}
 Q\cdot Q + Q \cdot a & = -2 \ , & k_1 &= 1\ , & k_2 &= 0 \ , & k_l &= 0 \ ,& Q\cdot J &=  1 > 0 \ .  
\end{align}
Furthermore, the central charges of the D1 branes CFT have the values
\begin{align}
c_L = 3 Q\cdot Q - 9 Q\cdot a + 2 &= 20 \ , & c_R & = 3 Q\cdot Q - 3 Q\cdot a = 6  \ .
\label{centralQ}
\end{align}
Finally, one can check that the unitarity constraint is satisfied 
\be
\sum_i \frac{k_i\,  \text{dim}\, G_i}{k_i + h_i^\vee} = \frac{N^2-1}{1+N} +1= N = 16 < c_L  \ .
\label{Kac}
\ee
We have added $1$ to account for the $U(1)$ factors of the gauge group. However, the direct computation of the central charges from the spectrum on the D1 branes yields different values for the central charges. The contributions of the two dimensional bosons ($1$ of $SO(1,1)$) and the MW fermions ($1/2_L$ and $1/2_R$ of $SO(1,1)$) are as follows
\be
\begin{array}{ccc}
\text{\bf $SO(1,1)$} & {\bf c_L }& {\bf c_R}\\
\hline\hline\\[-10pt]
1 & 1 & 1\\[2pt]
 1/2_L & 1/2 & 0\\[2pt]
 1/2_R & 0& 1/2\\[2pt] \hline
\end{array}
\label{central}
\ee
With the rules above one obtains the result
\begin{align}
c_L &= 4_{CM}+26+96_{D5}\ , & c_R & = 6_{CM} + 12 + 96_{D5} \ ,
\end{align}
where we have separated the contributions of the `center of mass' (CM) coordinates (coresponding to the adjoint hypermultiplet $(1,2,2,1)+(\frac12,2,1,2')_R$) and those of the non-chiral D1-D5 sector from the rest.  \\
The central charges computed from the D1 branes spectrum indeed do not match the ones computed with eq. \eqref{centralQ}. This is due to the fact that the D1 branes are on an orbifold fixed point and of the fact that D5 branes sit on the same orbifold fixed point. As we will see, the central charges for D1 branes are of the form
\be
c_L = c + 3 Q\cdot Q - 9 Q\cdot a +2  \geq  3 Q\cdot Q - 9 Q\cdot  a +2 \  , \quad  c_R  = c + 3 Q\cdot Q - 3 Q\cdot  a \geq  3 Q\cdot Q - 3 Q\cdot  a \ , \label{gencharges}
\ee
with $c\geq 0$ in general and equal to $6+96_{D5}$ in the example at hand.  
However, for D1 branes in the bulk central charges agree with eq. \eqref{centralQ}. Let us consider this case. The gauge group of the D1 branes in the bulk becomes orthogonal $SO(r)$. The massless spectrum is given in Table \ref{susy-bulk}.
\begin{table}[h!]
\centering
\begin{tabular}{c c}
{\bf Representation} & {\bf $SO(1,1)\times SU(2)_l \times SU(2)_R \times SO(4)$}\\
\hline\hline\\[-10pt]
$\frac{r(r-1)}{2}$ & $(0,1,1,1)+(\frac12,2,1,2)_L+(\frac12,1,2,2')_L$\\[2pt]
$\frac{r(r+1)}{2}$ & $(1,2,2,1)+(\frac12,1,2,2)_R$\\[2pt]
$\frac{r(r+1)}{2}$ & $(1,1,1,4)+(\frac12,2,1,2')_R$\\[2pt]
$r (n+\bar n)$ & $(\frac12,1,1,1)_L$\\[2pt]
 \hline
\end{tabular}
\caption{The spectrum of D1 branes displaced in the bulk on the $T^4/\mathbb{Z}_2$ supersymmetric orientifold. } \label{susy-bulk}
\end{table}

From the spectrum one finds the same anomaly polynomial as when the D1 branes were on top of an orbifold fixed point, that is
\be
I_4 = -r \left(\text{tr} R^2 - \frac12 \text{tr}F_1^2 \right)
\ee
and eqs. \eqref{chargeQ}-\eqref{Kac} still hold. However, the central charges computed from the spectrum of bulk D1 brane changes to the following values
\begin{align}
c_L & = 4_{CM} + 20\ , & c_R = 6_{CM}+6 \ ,
\end{align} 
which now match the charges in eq. \eqref{centralQ}, where the center of mass contributions have been removed. In our case these contributions come from the hypermultiplet in the symmetric representation corresponding to $(1,2,2,1)+(\frac12,1,2,2)_R$. 

 We considered a BPS D1 brane until now. One could wonder what would happen if one considers instead a non-BPS, but stable $\overline{D1}$ antibrane. The analysis above can be easily redone and 
the result is the expected one:  the corresponding charge vector is just the opposite of the one of D1 brane $Q_{\overline{D1}} = - Q_{D1}$. This could have been anticipated, since  the $\overline{D1}$ 
is an anti-instanton for the ${D5}$ background branes of the models, and therefore $Q=-b_2$. Clearly most constraints in eq. (\ref{constraints}) are violated in this case. In particular, the 2d chirality of all
fermions on the non-BPS string is flipped compared to the BPS case. One can write down appropriate constraints for such non-BPS defects, but they are in some sense mirrors of the BPS ones.

\subsection{A mini-landscape of models and string defects}

We consider here a general model with magnetized D9 branes and D5 branes satisfying the tadpole conditions on the supersymmetric $T^4/\mathbb{Z}_2$ orientifold \cite{Angelantonj:2000hi}. 
The gauge group is then of the form
\be
G=\prod_{\alpha} U(p_\alpha)_9 \times U(d)_5\ ,
\label{gengroup}
\ee
where, for simplicity, we consider a single stack of D5 branes sitting at the origin. The magnetic field on every D9 brane stack is chosen to be selfdual in order to preserve supersymmetry (see  \cite{Angelantonj:2000hi}). 
The RR tadpole conditions can be written as 
\be
\sum_{\alpha} (p_\alpha +\bar p_\alpha) n_1^{\alpha} n_2^{\alpha} = 32 \quad , \quad \sum_\alpha (p_\alpha + \bar p_\alpha) m_1^{\alpha} m_2^{\alpha} + d+\bar d = 32 \ , \label{rrtad}
\ee
whereas the twisted tadpoles are automatically satisfied by the Chan-Paton parametrization $R_{N_\alpha} = i(p_\alpha-\bar p_\alpha) = 0$ and $R_D = i(d - \bar d)=0$.
In the T-dual version, the D9/D5 branes correspond to intersecting D7 branes wrapping two-cycles inside $T^4/\mathbb{Z}_2$, whereas the O9 and O5 planes turn into O7 planes. 
The spectrum and consequently the anomaly polynomial in the intersecting brane version is conveniently described in terms of intersection numbers, for more details see Appendix D. 
In what follows one uses the intersecting D7 brane language for simplicity, although we will still talk about D9 and D5 branes, in line with our type I setup throughout the paper.

By making use of the spectra given in \cite{Angelantonj:2011hs}, one can show that the anomaly polynomial
has the general form
\begin{align}
I_8 &= \left(\text{tr} R^2 \right)^2 - \text{tr} R^2 \left[\frac18 \sum_\alpha I_{\alpha O}\,  \text{tr} F_\alpha^2 + \frac12 \text{tr} F_5^2 \right]+\frac1{16} \sum_{\alpha} I_{\alpha \alpha'} \left(\text{tr} F_\alpha^2 \right)^2\nonumber\\
& + \frac18\sum_\alpha \sum_{\beta \neq \alpha} (I_{\alpha \beta} + I_{\alpha \beta'})\,  \text{tr} F_\alpha^2 \, \text{tr} F_\beta^2 + \frac14 \sum_\alpha I_{\alpha 5}\, \text{tr} F_\alpha^2\, \text{tr} F_5^2 \ ,
\end{align}
where $F_\alpha$ is associated to the D9 brane stacks (including also the case of zero magnetization) and $F_{5}$ is associated to the D5 branes. Notice that the result above depends only on the toroidal intersection numbers (the conventions and definitions we are using are given in Appendix \ref{appD}). The absence of contributions from the orbifold fixed points can be traced to the fact that the branes are not fractional on this orientifold (in the T-dual picture, the two-cycle wrapped by the brane together with its image has zero twisted charges). From above, we can infer the anomaly lattice 
\begin{align}
a \cdot a &= 8\ , & a \cdot b_\alpha &= -\frac12  I_{\alpha O}\ , & a  \cdot b_5 & = -2\ , & b_\alpha^2 & = \frac12 I_{\alpha \alpha'}\ , \nonumber  \\
b_\alpha \cdot b_\beta &= \frac12(I_{\alpha \beta }+ I_{\alpha \beta'})\ , & b_5^2 & = 0\ , & b_\alpha \cdot b_5 & = I_{\alpha5}\ , \label{genlattice}
\end{align}
which is manifestly integral since $I_{\alpha \beta} + I_{\alpha \beta'}, I_{\alpha O}$ and $I_{\alpha \alpha'}$  are always even (see Appendix \ref{appD} for their definitions). The formula above suggests that one can identify the vectors $a$, $b_{\alpha}, b_5$ with the orientifold and brane two-cycles, respectively, wrapped by the O7/D7 in the T-dual description. Moreover, $\Omega$ is identified with the intersection form \cite{Kim:2019vuc} such that the anomaly lattice becomes a sublattice of $H_2(T^4/\mathbb{Z}_2)$.    
The most general (spacetime) D1 brane charges that can be obtained in the context of perturbative strings correspond to magnetized D5$'$ branes wrapping the whole $T^4/\mathbb{Z}_2$. In the intersecting branes picture, D1 and D5$'$ correspond to D3 branes wrapping a two-cycle in the torus orbifold. The boundary conditions for the various branes are summarized in Table \ref{d-branes}.
\begin{table}[h!]
\centering
\begin{tabular}{c|cccccc|cccc}
{\bf Brane} &0 & 1 & 2 & 3 & 4 & 5 & 6 & 7 & 8 & 9\\
\hline \hline\\[-10pt]
D9& $\times$ & $\times$ &$\times$ &$\times$ &$\times$ &$\times$ &$\times$ &$\times$ &$\times$ &$\times$\\[2 pt]
D5& $\times$ & $\times$ &$\times$ &$\times$ &$\times$ &$\times$ &$\bullet$ &$\bullet$ &$\bullet$ &$\bullet$\\[2 pt]
D5$'$& $\times$ & $\times$ &$\bullet$ &$\bullet$ &$\bullet$ &$\bullet$ &$\times$ &$\times$ &$\times$ &$\times$\\[2 pt]
D1& $\times$ & $\times$ &$\bullet$ &$\bullet$ &$\bullet$ &$\bullet$ &$\bullet$ &$\bullet$ &$\bullet$ &$\bullet$ \\[2 pt]
\hline
\end{tabular}
\caption{A cross denotes a direction parallel to the brane and a bullet denotes a direction orthogonal to the brane.}\label{d-branes}
\end{table}

In the following, we consider a magnetized D5$'$ brane (denoted D5$'_a$) supporting a $U(1)$ gauge group coupled to the general model of eq. \eqref{gengroup}. The magnetic field on the D5$'_a$ is assumed to be either selfdual or anti-selfdual for stability. Then, one can show explicitly from the D5$'_a$ spectra (see Appendix \ref{appD}) that the associated anomaly polynomial takes the form
\begin{align}
I_4 = \frac12 \left[-\frac12 I_{aO}\, \text{tr} R^2 + \frac12 \sum_{\beta} (I_{a \beta} + I_{a \beta'})\,  \text{tr} F_\beta^2 + I_{a5}\, \text{tr} F_5^2 + \frac12 I_{aa'}\, \chi(N) \right] \ .
\label{geninflow}
\end{align}
From above we can read the constraints that determine the charge vector $Q$. In this case we obtain
\begin{align}
Q \cdot a &= - \frac12 I_{aO}\ ,  & Q \cdot b_\beta & = \frac12(I_{a \beta} + I_{a \beta'})\ , & Q \cdot b_5 & = I_{a5}\ ,  & Q \cdot Q & = \frac12 I_{aa'} \label{genQ} \ .
\end{align}
The landscape of string-like charges that we obtain in this way is very large, since the D5$'$ magnetizations are only constrained by the stability (absence of tachyons) arguments. 
They however do not span all possible charges: for example using formulae from the Appendix \ref {appD} one can check $Q \cdot Q$ and $Q \cdot a$ are even. One could therefore question the completeness hypothesis \cite{Polchinski:2003bq}.  However, we showed in the previous subsection  by an explicit evaluation in the integral charge basis that these products are indeed even integers.  The completeness
hypothesis seem therefore to be satisfied. It can also be checked that only selfdual and anti-selfdual magnetic field configurations on the D5$'$ are stable (tachyon free). The self-dual configurations
are BPS, whereas the anti-selfdual are non-BPS.
 
Consider a D5$'_a$ having the same magnetization as a D9 stack (labeled by $\alpha$) such that it corresponds to a gauge instanton (the results apply also for the case of D1/D5 or D5$'$/D9 with zero magnetic field). In particular, we have for the wrapping numbers determining the magnetic field the identity $\otimes_i (m_i^\alpha, n_i^\alpha) =  \otimes_i (m_i^a, n_i^a) $. Then, after comparing eqs. \eqref{genlattice} and eqs. \eqref{genQ}, one can easily see that in this case the solution for the charge $Q$ is given by
\be
Q = b_\alpha
\ee
Conversely, one can use the brane instanton argument to argue that the equation above holds and then use the general anomaly lattice \eqref{genlattice} in order to derive the anomaly polynomial for D5$'_a$ branes in eq. \eqref{geninflow}. Finally, the constraints in \cite{Kim:2019vuc} are in principle satisfied if one chooses selfdual magnetic fields on the D5$'_a$. In particular, from above, we have $Q\cdot Q \geq 0$ and $Q \cdot b_\beta \geq 0$, $Q\cdot b_5 \geq 0$.  Using the Table 12 in the Appendix one can compute the central charges $c_L,c_R$ and compare with eqs. (\ref{gencharges}).  
Let us consider the particular case where all the multiplicities in Table 12 are positive. In addition, we take into consideration the possibility of having equal magnetization on the D5$'_a$ as for a stack of D9 branes with Chan-Paton charge denoted by $p_\alpha$. In the T-dual picture of intersecting branes it corresponds to a D3 brane wrapping the same two cycle as a D7 brane. In this case, using the spectra in Table 12, one finds
\begin{align}
 c_L &= 12 + 6 p_\alpha + \frac{3}{2} I_{a a'} + \frac{1}{2} I_{a O} +  \frac{1}{2} \sum_{\beta} (I_{a \beta} + I_{a \beta'}) p_\beta +  I_{a5} d = 4_{CM} + 8+6p_\alpha+ 3 Q \cdot Q - 9 Q \cdot a \ , \nonumber \\
 c_R &= 12 +6 p_\alpha+ \frac{3}{2} (I_{a a'} + I_{a O}) = 6_{CM}  + 6+6p_\alpha + 3 Q \cdot Q - 3 Q \cdot a \ .  \label{ml1}
\end{align}
where we used eqs. (\ref{genQ}) and  the identity
\be
\frac12\sum_{\beta} (I_{a \beta} + I_{a \beta'}) p_\beta +   I_{a5} d = 4   I_{a O}   \ ,  \label{ml2}
\ee
which is a consequence of the RR tadpole conditions eqs.  (\ref{rrtad}). One finds therefore in this case a formula of the type eqs. (\ref{gencharges}) with $c=6+6 p_\alpha$. One can show that the expression for $c$ is valid also in the case of D5$'$/D9 with zero magnetization or D1/D5 sitting on the same fixed point by replacing the Chan-Paton charge $p_\alpha$ with the one of the unmagnetized  D9 or D5 respectively. If the magnetization on the D5$'_a$ is different than the ones of the background branes then one simply sets $p_\alpha=0$, thus obtaining $c=6$ (this is for example the case for large magnetic fields on D5$'$ branes). 
The result  (\ref{ml1}) applies to a big ensemble of string defects with self-dual magnetic charges, but it is still not completely general. It does not apply for example to the cases where the multiplicity of bifundamental representations in Table 12 is negative.
We actually know from the previous Section the example of the bulk D1 branes for which $c=0$. However, $c=6$ seems to be the most generic value for arbitrarily large charges of BPS string defects, for which multiplicities in Table 12 are all positive. 

For general configurations, breaking supersymmetry, and in particular for the stable but non-BPS D5$'$  with anti-selfdual magnetic field configurations, the  
constraints do not hold anymore.  It would be interesting to find the set of contraints that charges of stable non-BPS string defects should satisfy for a consistent coupling to gravity.

\section{Six Dimensional Orientifold Models with Brane Supersymmetry Breaking}
\label{sec:bsb}

We now consider a $T^4/\mathbb{Z}_2$ with standard O9$_-$ planes and `exotic' O5$_+$ planes \cite{bsb}. The closed string spectrum is now modified with respect to the orientifold considered earlier (see Table \ref{bsb-closed}). 
\begin{table}[h!]
\centering
\begin{tabular}{ccc}
{\bf Multiplicity}& {\bf Multiplet} & {\bf Sector}\\
 \hline\hline\\[-10pt]
1 & Gravity & Untwisted\\[2pt]
1 & Tensor & Untwisted\\[2pt]
4 & Hypers & Untwisted\\[2pt]
16 & Tensors & Twisted\\[2pt] \hline
\end{tabular}
\caption{The closed string spectrum for the brane supersymmetry breaking $T^4/\mathbb{Z}_2$ orientifold.} \label{bsb-closed}
\end{table}
Notice that the closed string spectrum is supersymmetric. From the string theory realization, the cancellation of O9$_-$ and O5$_+$ tadpoles requires the introduction of 16 D9 branes and 16 $\overline{\text{D5}}$ (assumed to sit at the origin of the toroidal lattice). The presence of $\overline{\text{D5}}$ branes breaks supersymmetry, without introducing tachyons.  The question that we address in this section, in the spirit of the swampland program, is the following: can one understand if supersymmetry is broken or not for a 6d gauge theory coupled to the gravitational sector in Table \ref{bsb-closed}, from the consistency conditions of strings couplings to the tensors  in 6d? As one will see in what follows, supersymmetry breaking is manifest in the coupling to D1 branes of the 6d gauge theory in the perturbative string construction \cite{bsb}, with gauge group $SO(16)^2_9 \times USp(16)^2_5$. 

 The gauge group derived from the perturbative string construction has the form \cite{bsb}
 \be
G=SO(16)^2_9 \times USp(16)^2_5 \ .
\ee
The open string spectrum is given in Table \ref{bsb-open}. 
\begin{table}[h!]
\centering
\begin{tabular}{cc}
{\bf Field/Multiplet} & {\bf Representation}\\
\hline \hline\\[-10pt]
$A_\mu$ &  $(120,1;1,1)+(1,120;1,1)+(1,1;136,1)+(1,1;1,136)$\\[2pt]
$\chi_L$ &  $(120,1;1,1)+(1,120;1,1)+(1,1;120,1)+(1,1;1,120)$\\[2pt]
\hline\\[-10pt]
$(4 \phi, \psi_R)$ & $(16,16;1,1)+(1,1;16,16)$\\[2pt]
MW $\psi_L$ & $(16,1;16,1)+(1,16;1,16)$\\[2pt]
$2\phi$ & $(16,1;1,16)+(1,16;16,1)$ \\[2pt] \hline
\end{tabular}
\caption{The open string spectrum for the brane supersymmetry breaking $T^4/\mathbb{Z}_2$ orientifold.} \label{bsb-open}
\end{table}
\\
Notice that the bosons and fermions from the would be vector multiplet $(A_\mu, \chi_L)$ and the 1/2 `Hyper' $(2\phi, \psi_L)$ containing symplectic Majorana-Weyl spinors come in different representations with respect to the gauge group associated to the $\overline{\text{D5}}$ branes, breaking supersymmetry at the string scale. The appearance of {\it symplectic} Majorana-Weyl spinors (see \cite{Witten:1995gx} for a detailed discussion of their definition and consistency), which contain half of degrees of freedom compared to a standard 6d Weyl spinor, is possible due to the fact that the gauge group on the $\overline{\text{D5}}$-branes is symplectic. One could at first sight  build a supersymmetric field theory model with gauge group $G = SO(16)_9^2 \times SO(16)_5^2$ by putting the corresponding bosons and fermions in the same representations, since the resulting model would  have
exactly the same anomaly polynomial as the non-supersymmetric theory above, cancelling therefore all 6d gauge and gravitational anomalies. In string theory language, it would correspond to the introduction of D5 branes  (instead of $\overline{\text{D5}}$). However, even ignoring the fact that it would not satisfy the O5$_+$ tadpole conditions, it is not possible in 6d to define $1/2$ hypermultiplets without a symplectic gauge group (necessary for imposing the symplectic MW condition on the corresponding fermions). 

How does a string (D1 brane) coupled to this theory realize that supersymmetry is broken from the viewpoint of  consistency conditions in \cite{Kim:2019vuc}? 
In order to answer this question, we start from the anomaly polynomial. From the spectrum above, one can easily check the irreducible gravitational anomaly cancellation condition eq. \eqref{gravity} (with $N_T=17$). 
Since the number of tensor multiplets is bigger than the number of the non-abelian gauge factors, there are various ways to factorize the anomaly polynomial, related by $SO(1,17)$ transformations. However, only two 
factorizations are of interest for our purposes. The first is the geometrical one, in which each term in it corresponds to the coupling to the tensor fields defined by string perturbation theory. This form can be easily read off from the coupling of branes to the tensors, encoded for example in the partition functions. Since the anomaly lattice defined by $a \cdot a$, $a \cdot b_i$, $b_i \cdot b_j$ is independent of the chosen basis, this is enough to check that the anomaly lattice is integral, with the appropriate normalization
in Table 2. However, in the geometrical basis, the components of the vectors $a,b_i$ are generically not integers. It is nonetheless possible to switch to another basis that we call integral basis, 
less natural from string theory perspective, in which the components of vectors  $a,b_i$ are integers, proving the selfduality of the anomaly lattice.  In our example at hand, the geometrical basis for  the anomaly polynomial corresponds to the following factorized form 
\be
\begin{split}
I_8 &= \frac{1}{64} \left(\text{tr} F_1^2+ \text{tr} F_2^2 - \text{tr} F_3^2 - \text{tr} F_4^2 \right)^2 - \frac{1}{64} \left(-8\,  \text{tr} R^2+ \text{tr} F_1^2+ \text{tr} F_2^2 + \text{tr} F_3^2 + \text{tr} F_4^2  \right)^2 \\
&-\frac{1}{128} \left(\text{tr} F_1^2- \text{tr} F_2^2 + 4\, \text{tr} F_3^2 -4\, \text{tr} F_4^2  \right)^2 - \frac{15}{128} \left(\text{tr} F_1^2 - \text{tr} F_2^2 \right)^2 
\end{split} \ , \label{factor}
\ee
where $F_1,F_2$ ($F_3,F_4$) denote the gauge fields of the D9 branes  ($\overline{\text{D5}}$ antibranes). The first line in (\ref{factor})  is related to the couplings to the untwisted tensors and the second line 
to the couplings to the $16$ twisted tensors. The first term in the second line of (\ref{factor}) encodes the brane couplings to the twisted tensor at the fixed point where all $\overline{\text{D5}}$ are located,  whereas 
the second (last) term  contains the (equal) coupling to the other $15$ twisted tensors in the other fixed points, such that one has $N_T+1 = 18$ terms in total.

Analogously to its supersymmetric cousin,  the geometric interpretation is manifest in the string vacuum amplitudes (see \cite{Angelantonj:2002ct})
\begin{align}
&- {\tilde {\cal K}}_0 - {\tilde {\cal A}}_0 - {\tilde {\cal M}}_0 \sim \left[  (n_1+ {n}_2 - 32) \sqrt{v} -  \frac{ d_1+ {d}_2 - 32 }{\sqrt{v}}  \right]^2 C_4 C_4  +  \label{fac2} \\
& \left[  (n_1+ {n}_2 - 32) \sqrt{v} +   \frac{ d_1+ {d}_2 - 32 }{\sqrt{v}}  \right]^2 S_4 S_4   + \left[  (n_1-{n}_2 + 4 (d_1-{d}_2))^2 + 15  (n_1-{n}_2)^2 \right]  S_4 O_4 \ , \nonumber 
\end{align}
where the gauge group is parametrized here by $\left[SO(n_1) \times SO(n_2)\right]_9 \times  \left[SO(d_1) \times SO(d_2)\right]_{ 5}$ and $n_1=n_2=16$ and $d_1=d_2=16$ by the RR tadpole conditions.  
Unlike in the supersymmetric $T^4/\mathbb{Z}_2$ orientifold, the D-branes do couple to the sixteen anti-selfdual tensors (represented by the character $S_4 O_4$) from the twisted sector, therefore they are fractional branes. 

 From the geometrical factorization  (\ref{factor})  one can write (up to signs and permutations) the following representation for the vectors $a,b_i$ 
\begin{align}
a&=(0,-2\sqrt{2},0^{16})\ , & \Omega &= \text{diag} (1,-1^{17})\ , \label{ab1} \\
b_1& = \left(\frac{1}{\sqrt{2}}, \frac{1}{\sqrt{2}}, \frac12, \left(\frac12\right)^{15}\right)\ , & b_2 &= \left(\frac{1}{\sqrt{2}}, \frac{1}{\sqrt{2}}, -\frac12, \left(-\frac12\right)^{15}\right) \ , \label{ab2}\\
b_3& = \left(-\frac{1}{2\sqrt{2}}, \frac{1}{2\sqrt{2}}, 1, 0^{15}\right) \ , & b_4 &= \left(-\frac{1}{2\sqrt{2}}, \frac{1}{2\sqrt{2}}, -1, 0^{15}\right)\ .
\label{ab3}\end{align}
The corresponding products between the vectors $a$ and $b_i$ are then found to be
\begin{align}
a\cdot a& = -8 \ , & a\cdot b_i &= \left(\begin{array}{c}2 \\ 2\\ 1\\ 1 \end{array}\right)\ ,  & b_i \cdot b_j = \left(\begin{array}{cccc}
-4&4&-1&0\\
4&-4&0&-1\\
-1&0&-1&1\\
0&-1&1&-1
\end{array} \right) \ .
\label{bsbproducts}
\end{align}
Notice that we have integer entries for the products $b_i\cdot b_j$ only if one takes $\lambda_i =2$ for the orthogonal gauge group factors as indicated in Table \ref{lambda}. 

The requirement to find a basis with integral entries (see \cite{Seiberg:2011dr}) is also satisfied for the model under consideration. Indeed, one can check that the products in eq. \eqref{bsbproducts} can be obtained from the following vectors
\
\begin{align}
a&=(2,-2,0,0,0,-1^3,1,0,2,0^7) \ ,\label{bsbintegral1}\\
b_1 & = (0,1^4,0^{13})\ ,\label{bsbintegral2}\\
b_2 & = (2,-1^4,1,-1,1^2,0^9)\ ,\label{bsbintegral3}\\
b_3 & = (-1,1,0^4,1,0^{11})\ , \label{bsbintegral4}\\
b_4 & = (-1, 0^7,-1,0,-1,0^7)\ . \label{bsbintegral5}
\end{align}  
Let us check now if it is possible, in this case, to define a K\"{a}hler form $J=(J_0,J_1,\hdots,J_{17})$ that satisfies $J \cdot J>0$, $J\cdot a <0$ and $J\cdot b_i>0$. For the explicit vectors $a,b_i$ given in \eqref{ab1}-\eqref{ab3}, one finds the following system of inequalities 
\begin{align}
|J_0| &> |J_1|\ ,  & J_1&<0 \ , & J_0-J_1&>0 \ ,& J_0+J_1&<0 \ ,\label{JBSB}
\end{align}
which does not have a solution. Notice that if one relaxes the condition $J\cdot a <0$ it is still not possible to find a solution for $J$. We interpret this failure as a non-perturbative proof that {\it it is not possible to define a supersymmetric model} coresponding to this anomaly polynomial. Indeed, the would be supersymmetric model with $SO(16)_9^2 \times SO(16)_5^2$ gauge group does not exist in string theory or field theory. On the other hand we argue that a non-supersymmetric orientifold model is not required, in general, to satisy these conditions, since $J$ is related to supersymmetry. The impossibility to define $J$ does not depend on the choice of basis. Indeed, one can show the same result by making use of the integral basis in eqs. \eqref{bsbintegral1}-\eqref{bsbintegral5}. Two additional examples (with gauge groups $SO(16)_9 \times USp(16)_{5}$ and $[SO(8)^4]_9 \times [USp(8)^4]_{5}$) for which one cannot define $J$  are given in  Appendix \ref{appE}.

\subsection{D1 branes and the anomaly inflow}

If supersymmetry is broken in an arbitrary way, it is not clear if simple constraints can be formulated from coupling to string defects. However, if supersymmetry is broken locally on a collection of (anti-branes) as in our current example,  then it should be possible, at least if the string defects are geometrically separated from the source of supersymmetry breaking.  As one will see,  a detailed analysis suggests appropriate modifications to some constraints in case where supersymmetry is broken.

We start by considering the coupling to D1 branes of the brane supersymmetry breaking orientifold model. D1 branes at an orbifold fixed point are fractional
 (they have twisted charges), and their gauge group is of the form $SO(d_1) \times SO(d_2)$ in this case, where the two factors have opposite twisted charges. 
 Let us choose $d_1=d$ and $d_2=0$ such that we can write the total gauge group as
\be
SO(d)_1 \times \left[SO(n_1)\times SO(n_2)\right]_9 \times \left[USp(m_1)\times USp(m_2)\right]_5 \ . 
\ee
The vacuum amplitudes can be found in Appendix \ref{appC}. We give in Table \ref{bsb-fix} the spectrum of strings charged under the D1 branes, in case they sit at the same fixed point as the 
$\overline{\text{D5}}$ antibranes. 
\begin{table}[h!]
\centering
\begin{tabular}{c c}
{\bf Representation} & {\bf $SO(1,1)\times SU(2)_l \times SU(2)_R \times SO(4)$}\\
\hline\hline\\[-10pt]
 $\frac{d(d-1)}{2}$ & $(0,1,1,1)+(\frac12,1,2,2')_L$\\[2pt]
$\frac{d(d+1)}{2}$ & $(1,2,2,1)+(\frac12,2,1,2')_R$\\[2pt]
$d n_1$ & $ (\frac12,1,1,1)_L$\\[2pt]
$d m_1$ &$(\frac12,1,1,2')_L$\\[2pt]
$d m_2$ & $(\frac12,1,1,2)_R$\\[2pt]\hline
\end{tabular}
\caption{The spectrum of D1 branes at a fixed point (same as the $\overline{\text{D5}}$ antibranes) on the $T^4/\mathbb{Z}_2$ brane supersymmetry breaking orientifold.} \label{bsb-fix}
\end{table}

From the spectrum, one can easily see that the anomaly polynomial corresponding to the D1 branes can be written in the following form
\be
I_4 =- \frac{d}{2} \left(\text{tr} R^2 - \frac12 \text{tr} F_1^2 - \text{tr} F_3^2 + \text{tr} F_4^2+d \chi(N) \right)\ .
\ee
After comparing with the general form in eq. \eqref{polyQ2} we can infer the following conditions for the charges
\begin{align}
Q \cdot Q & =-1\ , & Q \cdot a & =-1 \ , & Q \cdot b_1 & =1 \ , & Q \cdot b_2 &=0\ , & Q \cdot b_3 & = 1\ , & Q \cdot b_4 & =-1 \ , 
\label{Qsystem}
\end{align}
where we have considered the minimal Chan-Paton $d=1$. The solution for the vector of charges $Q$, for the basis in eqs. \eqref{ab1}-\eqref{ab3} is then found to be
\begin{align}
Q &= \left(\frac{1}{2\sqrt{2}}, -\frac{1}{2\sqrt{2}},-1, 0^{15} \right) \ .
\end{align}
Notice that the position of the $-1$ value can be permuted by placing the D1 branes on a different orbifold fixed point. Let us consider the following two choices of charge vectors
\begin{align}
Q &= \left(\frac{1}{2\sqrt{2}}, -\frac{1}{2\sqrt{2}}, -1, 0^{15} \right)\ , & \tilde Q & = \left(\frac{1}{2\sqrt{2}}, -\frac{1}{2\sqrt{2}}, 0, -1, 0^{14} \right) \ ,
\end{align}
which reflect placing the  branes on different fixed points.  It is interesting to notice that the D1 brane is an anti-instanton for the $\overline{\text{D5}}$ when they sit on the same fixed point. Then, from above, we see that indeed $Q=-b_3$ is the solution of the inflow. The products of the two charges above with the vectors in eqs. \eqref{ab1}-\eqref{ab3} yield 
\begin{align}
k_1 = Q \cdot b_1 &= 1\ , & k_2 = Q\cdot b_2 &=0\ , & k_3  = Q\cdot b_3 &=1\ ,  & k_4 = Q\cdot b_4 & =-1 \ , \\
k_1 = \tilde Q \cdot b_1 &=1\ , & k_2 = \tilde Q\cdot b_2 &=0\ , & k_3  = \tilde Q\cdot b_3 &=0 \ , & k_4 =\tilde  Q\cdot b_4 & =0 \ . 
\end{align}
Notice  the $k_4$ violation of the positivity requirement, in the first line above. This is due  to the fact that the Kac-Moody algebra in this case is realized on both left and right sectors. Indeed, due to the presence of $\overline{\text{D5}}$ on the same fixed point as the D1 branes, this gives rise to right-handed fermions in the spectrum (see Table \ref{bsb-fix}). Also, $k_3$ is positive as the corresponding spectrum consists of left-handed fermions. If one places the D1 branes on a different fixed point (corresponding to the charge $\tilde Q$) then strings stretched between the D1 and $\overline{\text{D5}}$ become massive and thus they cannot detect  supersymmetry breaking (in the infrared). 
From eq. \eqref{Qsystem} we also have $k_l=0$. The products $Q \cdot Q$ and $Q \cdot a$ give the same result for both choices $Q$ and $\tilde Q$ (as in eq. \eqref{Qsystem}) and thus the minimal central charges $c_L$, $c_R$ are the same
\begin{align}
c_L  &=8\ , & c_R &  = 0 \ .
\end{align}
Finally, it turns out that the unitarity constraint is saturated in this case
\be
\sum_i \frac{k_i\, \text{dim}\, G_i}{k_i + h_i^\vee} =  \frac{N(N-1)}{2} \frac{1}{1 + N-2} = \frac{N}{2} =8 =c_L \ .
\ee
It is useful to estimate the central charges from the D1 brane spectrum. By making use of the rules in eq. \eqref{central} one obtains
\begin{align}
c_L &= 4_{CM} + 8 + 16_{D5} \ , \\
c_R &  = 6_{CM} + 0+ 16_{D5}\ .
\end{align}
Notice that if one displaces the D1 branes at a different fixed point then the $16_{D5}$ contribution disappears and the central charges coincide with the minimal ones computed from $Q$ and $a$. In general, the central charges match the form in eq. \eqref{gencharges}. A vector of integer tensor charges can be found in the integral basis  \eqref{bsbintegral1}-\eqref{bsbintegral5},  after solving the constraints \eqref{Qsystem} with the corresponding $a$ and $b_i$. In this case one finds
\be
Q = (1,-1,0^4,-1,0^{11}) \ .
\ee
As in the supersymmetric case, we now consider D1 brane probes in the bulk for which the central charges match precisely the ones computed from $Q$ and $a$. The massless spectrum of the bulk D1 branes is given in Table \ref{bsb-bulk}.
\begin{table}[h!]
\centering
\begin{tabular}{c c}
{\bf Representation} & {\bf $SO(1,1)\times SU(2)_l \times SU(2)_R \times SO(4)$}\\
\hline\hline\\[-10pt]
$\frac{d(d-1)}{2}$ & $(0,1,1,1)+(\frac12,2,1,2)_L+(\frac12,1,2,2')_L$\\[2pt]
$\frac{d(d+1)}{2}$ & $(1,2,2,1)+(\frac12,1,2,2)_R$\\[2pt]
$\frac{d(d+1)}{2}$ & $(1,1,1,4)+(\frac12,2,1,2')_R$\\[2pt]
$d (n_1+n_2)$ & $(\frac12,1,1,1)_L$\\[2pt]
 \hline
\end{tabular}
\caption{The spectrum of D1 branes displaced in the bulk on the $T^4/\mathbb{Z}_2$ brane supersymmetry breaking orientifold.} \label{bsb-bulk}
\end{table}

From above, one finds that the anomaly polynomial for the D1 branes displaced in the bulk becomes
\be
I_4 = -d \left(\text{tr} R^2 - \frac14 \text{tr} F_1^2 - \frac14 \text{tr} F_2^2 \right) \ .
\ee
After matching the polynomial above with the general form in eq. \eqref{polyQ2}, one finds the following constraints for the charges
\begin{align}
Q \cdot Q &= 0\ ,  & Q \cdot a & = -2 \ ,  & Q \cdot b_1 & =1\ ,  & Q \cdot b_2 & =1\ , & Q \cdot b_3 & = 0\ ,  & Q \cdot b_4 & =0 \ . \label{Qsystem2}
\end{align} 
Notice that even in this case with localized supersymmetry breaking,  we are able to identify null charged strings satisfying $Q \cdot Q=0$, for bulk D1 branes, which are away also away 
from the supersymmetry breaking source. 
 In the basis given in eqs. \eqref{ab1}-\eqref{ab3},  the equations above determine the tensor charges of the D1 branes as follows
\begin{align}
Q &= \left(-\frac{1}{2\sqrt{2}}, \frac{1}{2\sqrt{2}},0^{16} \right) \ .
\end{align}
Notice that from eq. \eqref{Qsystem2} follows that we have $k_1=k_2=1$ integral only if we take $\lambda=2$ for the orthogonal gauge group factors (with $\lambda=1$ they are equal to $1/2$). In the integral basis of eqs. \eqref{bsbintegral1}-\eqref{bsbintegral5} the charge $Q$ is given by
\be
Q = (2,-1,0^4,-1,0,1,0,1,0^7) \ .
\ee 
The central charges of the D1 brane CFT are given by
\begin{align}
c_L = 3 Q \cdot Q-9 Q \cdot a + 2 &= 20 \ ,  & c_R = 3 Q\cdot Q- 3 Q \cdot a &= 6 \ .
\end{align}
We also have $k_l =0 $. Let us now check the central charges from the massless spectrum of bulk D1 branes. It easy to see that we have
\begin{align}
c_L & = 4_{CM}+20\ , & c_R & = 6_{CM}+6 \ ,
\end{align}
where we have separated the contributions of the center of mass coordinates corresponding to the hypermultiplet $(1,2,2,1)+(\frac12,1,2,2)_R$ in the symmetric representation of the D1 brane gauge group.

Similarly to the SUSY example, one could considers instead a  stable $\overline{D1}$ antibrane. The corresponding charge vector is again just the opposite of the one of a D1 brane $Q_{\overline{D1}} = - Q_{D1}$.  Indeed,  the $\overline{D1}$  is an instanton for the $\overline{\text{D5}}$ background antibranes of the models, and therefore $Q=b_3$.  Most constraints in 
 eq. (\ref{constraints}) are violated in this case and the 2d chirality of all fermions on the $\overline{D1}$ is flipped compared to the D1  case.

\subsection{A mini-landscape of models and string defects}

As for the supersymmetric case, we consider now a general model with magnetized D9 branes and $\overline{\text{D5}}$ branes for the $T^4/\mathbb{Z}_2$ brane supersymmetry breaking orientifold  \cite{Angelantonj:2000hi}. One has to distinguish, in this case, between the D9 with zero magnetization having an orthogonal gauge group and the magnetized D9's with unitary gauge group. The gauge group is then of the form 
\be
[SO(n_1)\times SO(n_2)]_9 \times [USp(m_1) \times USp(m_2)]_{ 5} \times \prod_{\alpha} U(p_\alpha) \ .
\ee
Notice that one has to choose antiselfdual magnetic fields on the D9 branes such that supersymmetry is broken without generating tachyons.  
The RR tadpole conditions are
\begin{equation}
\begin{split}
&\sum_\alpha (p_\alpha+\bar p_\alpha )n_1^\alpha n_2^\alpha +n_1+n_2=32 \ , \\
&\sum_\alpha (p_\alpha+\bar p_\alpha ) m_1^\alpha m_2^\alpha +m_1+m_2=-32 \ , \\
&\sum_\alpha (p_\alpha+\bar p_\alpha)\epsilon_{ij}^\alpha+R_N+R_D=0 \ . 
\end{split} \label{mlbsb1}
\end{equation}
In   (\ref{mlbsb1}),  $\epsilon_{ij}^\alpha$ are twisted charges equal to $\pm 1$ if the brane stack $\alpha$ passes through the fixed point
labeled by  $ij$ and zero otherwise (i,j=1,...,4).  In addition, the orbifold actions on the Chan-Paton factors
$R_N,R_D$ are different from zero only  for the D9/D5 branes which pass through the corresponding orbifold fixed point.
The conventions and definitions we are using are given in Appendix \ref{appD}. 

By making use of the spectra given in \cite{Angelantonj:2011hs}, one can show that the anomaly polynomial
has the general form \footnote{See Appendix \ref{appD} for the definitions of the various quantities.}
\begin{align}
I_8 &= - \left(\text{tr} R^2 \right)^2 - \text{tr} R^2 \left(\frac18 \sum_\alpha \tilde I_{\alpha O}\, \text{tr} F_\alpha^2 - \frac14 \sum_{i=1}^4 \text{tr} F_i^2\right)+ \frac18 \left [ \sum_\alpha \frac12 (I_{\alpha \alpha'} -8) \left(\text{tr} F_\alpha^2 \right)^2 - \sum_{i=1}^4 \left(\text{tr} F_i^2 \right)^2 \right ]\nonumber\\
&+ \frac18\sum_\alpha \sum_{\beta \neq \alpha} (I_{\alpha \beta}+I_{\alpha \beta'}-2 \epsilon_\alpha \epsilon_\beta S_{\alpha \beta}) \text{tr}F_\alpha^2\, \text{tr} F_\beta^2 + \frac18\sum_{\alpha} (I_{\alpha 9} - \epsilon_\alpha S_{\alpha 9}) \text{tr} F_\alpha^2\,  \text{tr} F_1^2\nonumber\\
&+ \frac18\sum_{\alpha} (I_{\alpha 9} + \epsilon_\alpha S_{\alpha 9}) \text{tr} F_\alpha^2\,  \text{tr} F_2^2 -  \frac18\sum_{\alpha} (I_{\alpha 5} + \epsilon_\alpha S_{\alpha 5}) \text{tr} F_\alpha^2\,  \text{tr} F_3^2 -   \frac18\sum_{\alpha} (I_{\alpha 5} - \epsilon_\alpha S_{\alpha 5}) \text{tr} F_\alpha^2\,  \text{tr} F_4^2\nonumber\\
&+\frac14 \left(\text{tr} F_1^2\, \text{tr} F_2^2 +  \text{tr} F_3^2\, \text{tr} F_4^2 \right) - \frac18\left(\text{tr} F_1^2\, \text{tr} F_3^2+\text{tr} F_2^2\, \text{tr} F_4^2   \right) \ ,
\end{align}
where $F_1,F_2$ are associated to the unmagnetized D9 branes with orthogonal gauge group, $F_3,F_4$ are associated to the $\overline{\text{D5}}$ branes with symplectic gauge group and $F_\alpha$ to the magnetized D9 branes with unitary gauge group. Notice that now one receives contributions from the orbifold fixed points signaling the fact that the branes are fractional. We can now read the anomaly lattice from above
\begin{align}
a \cdot b_\alpha &= - \frac12 \tilde I_{\alpha O}\ , & b_\alpha^2 &= \frac12(I_{\alpha \alpha'}-8)\ ,  & b_\alpha \cdot b_\beta = \frac12(I_{\alpha \beta}+I_{\alpha \beta'}-2 \epsilon_\alpha \epsilon_\beta S_{\alpha \beta}) \ , \nonumber\\
b_\alpha \cdot b_1 & = I_{\alpha 9}- \epsilon_\alpha S_{\alpha 9}\ , & b_\alpha \cdot b_2 & = I_{\alpha 9}+ \epsilon_\alpha S_{\alpha 9} \ , \nonumber\\
 b_\alpha \cdot b_3 & = -\frac12(I_{\alpha 5}+ \epsilon_\alpha S_{\alpha 5})\ ,  & b_\alpha \cdot b_4 & =- \frac12(I_{\alpha 5}- \epsilon_\alpha S_{\alpha 5})\ ,
\label{bsblattice}
\end{align}
supplemented with eq. \eqref{bsbproducts} for the unmagnetized part. It is easy to see that the anomaly lattice is manifestly integral.  Similarly, as for the supersymmetric case, in the intersecting brane picture, the vectors $a$, $b_{\alpha}$, $b_i$ can be identified with the two-cycles wrapped by the orientifold planes and the branes and $\Omega$ can be identified with the intersection form \cite{Kim:2019vuc} such that the anomaly lattice becomes a sublattice of $H_2(T^4/\mathbb{Z}_2)$ (involving also the twisted part in this case). 

We consider the coupling to a magnetized D5$'_a$ brane with $U(1)$ gauge group (thus assuming non-zero magnetic field\footnote{The case with zero magnetic field has to be considered separately as the gauge group is not unitary. One can infer the results in this case by using the fact that the unmagnetized D5$'$ is an instanton for the unmagnetized D9.}). Furthermore, the magnetic field on the D5$'_a$ is assumed to be either self-dual or anti-selfdual. One can show from the massless spectra for D5$'_a$ branes (see Appendix \ref{appD}) that the anomaly polynomial has the form
\begin{align}
I_4 &= \frac12 \left[-\frac12 \tilde I_{aO}\, \text{tr} R^2 + \frac12 \sum_{\beta} (I_{a \beta} + I_{a \beta'}-2 \epsilon_a \epsilon_\beta S_{a\beta})\, \text{tr} F_\beta^2 +\frac12(I_{a9}-\epsilon_a S_{a9}) \text{tr} F_1^2 + \frac12 (I_{a9}+\epsilon_a S_{a9}) \text{tr} F_2^2\right. \nonumber\\
 &\left. -\frac12(I_{a5}+\epsilon_a S_{a5}) \text{tr} F_3^2 - \frac12(I_{a5}-\epsilon_a S_{a5}) \text{tr} F_4^2+\frac12 (I_{aa'}-8)\, \chi(N) \right]
\end{align}
From above we can read the constraints that determine the charge vector $Q$. In this case we obtain
\begin{align}
Q \cdot a &= - \frac12 \tilde I_{aO} & Q \cdot b_\beta & = \frac12(I_{a \beta} + I_{a \beta'}-2 \epsilon_a \epsilon_\beta S_{a \beta})\nonumber \\
 Q \cdot b_1 & = I_{a9}-\epsilon_a S_{a9} & Q \cdot b_2 & = I_{a9}+\epsilon_a S_{a9} \nonumber\\
  Q \cdot b_3 & = -\frac12(I_{a5}+\epsilon_a S_{a5}) & Q \cdot b_4 & = -\frac12(I_{a5}-\epsilon_a S_{a9}) \nonumber\\
Q \cdot Q & = \frac12 (I_{aa'}-8)\label{bsbgenQ}
\end{align}
Consider a D5$'_a$-brane having the same {\it non-zero} magnetization as a D9 brane such that it corresponds to a gauge instanton. Then, after comparing eq. \eqref{bsblattice} and eq.\eqref{bsbgenQ}, one can easily see that in this case the solution for the charge $Q$ is again given by 
\be
Q = b_\alpha \ .
\ee
For the case of unmagnetized D5$'$ one finds $Q=b_1/2$ or $Q=b_2/2$, the factor of $2$ arising from taking $\lambda_{SO}=2$. 

As in the SUSY example, the landscape of string-like charges that we obtain in this way is very large, since the D5$'$ magnetizations are only constrained by the stability (absence of tachyons) arguments. 
It can be checked that only selfdual and anti-selfdual magnetic field configurations on the D5$'$ are stable (tachyon free).  However,  they do not span all possible charges:  one can again check that $Q \cdot Q$ is even. In this case, we didn't check the compatibility with the completeness hypothesis \cite{Polchinski:2003bq}. Finally,  in this case we expect most of the constraints in \cite{Kim:2019vuc} to be violated. For example, for antiselfdual magnetic fields on D5$'$ we find that $Q \cdot Q <-1$ is possible and generic.  

In conclusion, both the SUSY and the BSB example contain a large class of non-BPS but stable string defects.  It would be interesting to find new consistency constraints on the charges on non-BPS string 
defects coming from anomaly inflow and unitarity arguments. 

\subsection{Morrison-Vafa F-theory $SO(8)^8$ example}

Interestingly, there is  an F-theory realization with the same closed string sector as the $T^4/\mathbb{Z}_2$ orientifold with O9$_-$ and O5$_+$ planes and thus having $N_T =17$,  with pure super Yang-Mills sector 
and gauge group $SO(8)^8$ \cite{Morrison:1996pp}, \cite{Blum:1996hs}. The model has no perturbative orientifold realization, due to the impossibility to cancel the RR tadpole in a supersymmetric way, as discussed in the brane supersymmetry breaking (BSB) example before. The anomaly polynomial can be written in one factorization as
\be
\begin{split}
I_8 &= \frac{1}{64}  \left( \text{tr} F_1^2 + \hdots +\text{tr} F_8^2 \right)^2 - \frac{1}{64}  \left(-8\, \text{tr} R^2+ \text{tr} F_1^2 + \hdots +\text{tr} F_8^2 \right)^2\\
&-\frac{1}{16} \left[2\left(\text{tr}F_1^2 \right)^2+\hdots +2 \left(\text{tr}F_8^2\right)^2 \right]
\end{split} \ ,
\ee
From above, the products between the vectors $a,b_i$ associated to the polynomial can be found to be
\begin{align}
a\cdot a &= -8\ ,  & a \cdot b_i & =2 \ , & b_i \cdot b_j & = - 4 \delta_{ij} \ .
\end{align}
A choice for $\Omega$ and $a,b_i$ consistent with the products above is the following
\begin{align}
\Omega& = \text{diag} (1,-1^{17})\ , &\ & & a & = (-3,1^{17}) \ , \\
  b_1 & = (0,-2,0^7,0^9) \ , &\dots \ , && b_8 &=(0,0^7,-2,0^9) \ . 
\end{align}
The constraints for finding a  K\"{a}hler form $J$ are then
\begin{align}
J \cdot J &= J_0^2 - \vec J^2 >0\ , & J \cdot b_i &= 2 J_i >0 \ , & J\cdot a = -3 J_0 - \sum_{I=1}^{17} J_I <0 \ .
\end{align}
One can check that a possible solution is
\be
J = (3, 1^8,0^9) \ .
\ee
Let us again check that the model, being a F-theory model, is not ruled out by the techniques of \cite{Kim:2019vuc}, using the vectors defined just above. We consider a generic string which couples to the tensor fields with charge $Q=(q_0,q_1,...,q_{17})\in\mathbb{Z}^{18}$, subject to the consistency conditions the first line of \eqref{constraints}. We have $0\leq k_{i=1,...,8}=Q\cdot b_i=2 q_i$.  To find
non-trivial constraints, we should  consider at least one non-zero $k_i$, meaning one non-zero $q_{i=1,...,8}$. Thus from $Q\cdot Q\geq -1$, we derive that $q_0^2\geq\sum_{i=9}^{17}q_i^2$. From $Q\cdot J\geq 0$ we find that $3q_0\geq \sum_{i=1}^{8}q_i\geq 0$, and it follows that
\be
3q_0\geq\sqrt{9\sum_{i=9}^{17}q_i^2}= \sqrt{\sum_{i=9}^{17}q_i^2+\sum_{i=9}^{17}\sum_{j>i}^{17}(q_i^2+q_j^2)}\geq \sum_{i=9}^{17}|q_i| \ ,
\ee
where for the last inequality one uses $q_i^2+q_j^2 \geq \pm 2 q_i q_j$. From above one further has that $Q\cdot a=-3q_0-\sum_i q_i\leq -\sum_{i=1}^8 q_i$. Then, 
\be
c_L=3 Q\cdot Q-9Q\cdot a +2\geq 9 \sum_{i=1}^8 q_i -1 \ ,
\ee
and
\be
\sum_{i=1}^8\frac{k_i\,\text{dim }G_i}{k_i+h^\vee_i}=\sum_{i=1}^8\frac{28 q_i}{q_i+3}\leq \sum_{i=1}^8 7 q_i\leq 9 \sum_{i=1}^8 q_i -1\leq c_L \ .
\ee
Thus, we conclude that the model passes the consistency test of \cite{Kim:2019vuc}.

Notice that the model is compatible with the null charged strings condition, proposed in Section  \ref{sec:null}. 
Indeed, a null charge  vector in this case can be chosen to be
\be
Q = (3,1,0^8,-1^8) 
\ee
giving rise to the following data
\begin{align}
Q \cdot Q & = 0\ , & Q \cdot a &= -2\ , &  \\
 Q\cdot J &= 8>0\ , & \vec k &= (2,0^7) \ , \\
c_L & = 20 \ , & c_R & = 6\ , & k_l =0 \ .
\end{align}
It can also easily be checked that the last condition in (\ref{constraints}) is verified.  

One could imagine that the BSB vacuum discussed in Section \ref{sec:bsb} and the F-theory supersymmetric vacuum discussed above are related to each other in some way, being coupled to the same gravitational spectrum. However, we are not aware of an obvious connection between the two. The perturbative BSB construction can be deformed in various ways by moving or recombining branes, see \cite{Angelantonj:2011hs} and
our Appendix.  The $SO(8)^8$ F-theory model on the other hand has no possible deformations parameters in six dimensions and is isolated in the space of vacua. 
No higgsing phenomenon or brane recombination seems to account for an eventual transition between the BSB vacuum, of gauge group $SO(16)^2 \times USp(16)^2$, to the supersymmetric F-theory one with
smaller gauge group $SO(8)^8$. The fact that the latter has no perturbative orientifold realization hints towards a nonperturbative transition (if any), of unknown nature to us. 



\section{Summary of  Results and Conclusions}

Our main motivation for this paper was to investigate consistency constraints for quantum gravity models coming from their coupling to string defects, in settings with minimal supersymmetry in 6d, or broken (non-linear)
supersymmetry  localized on antibranes of the brane supersymmetry breaking (BSB) type, with a supersymmetric gravitational (closed string) spectrum. 
We confronted consistency conditions on 6d theories from anomaly inflow  and conformal symmetry constraints derived in \cite{Kim:2019vuc}   with perturbative
6d orientifold constructions with minimal supersymmetry and  BSB models,  by adding  string defects with a large set of charges. 

 In all cases we studied and other string constructions existing in the literature that we were able to check, we found that string defects having null charges $Q \cdot Q =0$ do exist. Their existence is guaranteed in geometric string compactifications, and could therefore be a string lamppost signal. Based on this argument and our scan of the landscape we conjectured their existence in any quantum gravity theory.
 By investigating their presence, we were able to exclude some 6d models with no current string or F-theory realization, otherwise consistent with the other constraints. Furthermore, if the null strings hypothesis is violated in string theory, we expect it to correspond to truly non-geometric compactifications.

In our orbifold examples, we found generic constraints on the string charges coming from the geometry of the models: the left and right central charges on D1 strings have additional contributions compared to \cite{Kim:2019vuc},  if they intersect D5 branes leading to massless fermions, or if the D1 strings have non-trivial Chan-Paton (CP) factors. The latter can happen even for the minimum CP factor. In all cases, if D1 strings are not at orbifold singularities their central charges fit precisely with the formulae in \cite{Kim:2019vuc}. 
All these results hold also in BSB vacua, if the D1 strings are geometrically separated from the supersymmetry breaking (antibrane) source.
 
 We also analyzed a large class of string defects coming from D5 branes (called D5$'$ in the text) wrapping the four orbifolded dimensions, with self-dual and anti-selfdual magnetic fluxes in their worldvolume. 
 In the SUSY vacua, the self-dual magnetized D5$'$ are BPS, whereas the anti-selfdual ones are non-BPS but stable. In the BSB case, both of them are non-BPS but stable. Interestingly, their charges do not span
 all possible set of integers, however in the SUSY case we checked the compatibility with the completeness hypothesis. For the BSB non-SUSY case, we didn't perform a complete check of all  consistent charges
 from unitarity arguments and therefore we didn't check the validity of the completeness principle. 
  
One important result is the proof that the anomaly polynomial and the constraints in \cite{Kim:2019vuc} applied
to the supergravities of the  6d BSB type \cite{bsb} and similar constructions, `know'  that the 6d gauge theory derived from the perturbative type I spectrum breaks supersymmetry (realizes it nonlinearly).  More precisely, in the brane supersymmetry breaking case, the anomaly polynomial shows that generically one cannot define a Kahler form $J$, which clearly indicates that its existence is tied to supersymmetry.  
In addition, the flavor central charges for bulk gauge fields living on antibranes ($\overline{\text{D5}}$ in our examples) can be negative, since the fermions at the intersection of the strings with the antibranes have
opposite chirality compared to the supersymmetric case. 
 This also shows that some  models which do not fulfil the constrains of \cite{Kim:2019vuc}-\cite{Katz:2020ewz}, could have non-supersymmetric solutions, i.e. the gravitational sector (closed sector in string constructions) could still lead to a consistent theory, but with non-linear supersymmetry in the gauge/brane sector. However, the isolated F-theory supersymmetric vacuum with the same gravitational sector \cite{Morrison:1996pp} 
does satisfy the constraints coming from coupling to string defects. The unknown transition between the non-supersymmetric perturbative orientifold vacuum and the supersymmetric rigid F-theory one remains mysterious
to us.

A more technical summary of our results is: 
\begin{itemize}

\item
If $N_T \geq 1$, D1 strings in the bulk (away from orbifold fixed points) couple only to the untwisted tensor multiplet, which is split as usual into a self-dual and an anti self-dual components. 
The charge vector for all such D1 strings satisfy $Q \cdot Q=0$. Such strings\footnote{Regular D1 strings at orbifold fixed points can also have null charges, since they have no twisted charges.} 
should exist in any model with tensor
multiplets coming from a geometric string compactification.  It is tempting to contemplate the conjecture that null charged strings should always exist in a consistent 6d theory coupled to strings. 
We gave examples of 6d theories satisfying all other consistency conditions (\ref{constraints}), that would be excluded by this new condition. 

\item
The central charges of the BPS string defects (D1 brane or D5 branes wrapped over four internal dimensions, called  $D5'$ in the main text, with self-dual magnetic configurations),  
are generally of the form 
\begin{equation}
c_L = c + 3 Q\cdot Q - 9 Q\cdot a +2  \geq  3 Q\cdot Q - 9 Q\cdot a +2 \quad  , \quad c_R = c + 3 Q\cdot Q - 3 Q\cdot a \geq  3 Q\cdot Q - 3 Q\cdot a \ , 
\end{equation}
where $c \geq 0$ is a left-right symmetric contribution to the central charge due to  either vector-like massless  degrees of freedom at the intersection of D1 and D5 branes or the $D5'$ and the D9 branes, 
or to the additional Chan-Paton charges living on the string. This result holds also if supersymmetry is broken in a localized way in the internal space, \`a la BSB for branes far from the supersymmetry breaking source\footnote{In our BSB example,
also for D1 branes sitting at the fixed point where supersymmetry is broken by antibranes. But not for most general non-BPS stable string defects, coming from D5 branes wrapping the four orbifolded dimensions. }. 

\item
Non-BPS string defects (D5 wrapped over four internal dimensions, with anti-self dual magnetic configurations in SUSY vacua, or both self-dual and anti-self-dual magnetic fields in BSB vacua)
violate  generically all constraints, eqs. (\ref{constraints}), except the last one.

\item The Kahler form $J$ cannot be in general be defined in brane supersymmetry breaking models. This means that its existence and properties is intrinsic to supersymmetric models. 

\item
Flavor charges $k_i = Q\cdot b_i$ on D1 branes for bulk (six dimensional) gauge symmetries are non-negative in supersymmetric models, since the fermionic zero-modes at the intersection between the bulk branes and the D1 branes (string) are left-handed. For brane supersymmetry breaking models containing antibranes, if they intersect the D1 branes, they lead to fermionic zero modes which can have opposite chiralities  and therefore 
to negative $k_i$ in a consistent way.  Moving the D1 strings away from the antibranes render these states massive, and as a result the flavor charges become positive.  
For the wrapped D5$'$ string defects landscape, $k_i$ are positive in the self-dual magnetic configurations leading to BPS string defects, whereas they are typically negative for the non-BPS string defects, due to the flip of 2d chirality of fermions. 

\end{itemize}

Let us comment that for the spontaneous supersymmetry breaking (for example \`a la Scherk-Schwarz  \cite{Scherk:1979zr}, \cite{Rohm:1983aq}), the consistency conditions from couplings to defect strings  are the same as in the supersymmetric case. The reason is that due to its spontaneous nature, supersymmetry breaking is adiabatic and is restored in the decompactification limit.  Since the consistency conditions should hold for all values of the adiabatic parameter, it holds in particular in the supersymmetric limit.  This is the case for example for the recent constructions with suppressed one-loop vacuum energy \cite{Abel:2015oxa}.  
For tachyon-free models with no supersymmetry at all, we expect that the constraints discussed in  this paper do not apply.  This could be the case for the 10d $SO(16) \times SO(16)$ heterotic string \cite{so(16)xso(16)} (however, its continuous interpolation with the superstring could make it similar to the Scherk-Schwarz examples), the 0'B orientifold \cite{O'B} 
and their compactifications \cite{Blaszczyk:2014qoa}, \cite{Angelantonj:1998gj}.    

It would be very interesting to investigate along these lines four-dimensional field theory models with minimal supersymmetry or no supersymmetry, with tensor fields, dual to axions in four dimensions, coupling
to string defects. Nontrivial restrictions would impose new swampland constraints on physics beyond the Standard Model, coming from inflow arguments.  


\section*{Acknowledgments}

We thank Gianfranco Pradisi, Augusto Sagnotti and Washington Taylor for useful discussions and correspondence. 
C.A. and C.C. would like to thank CPHT at Ecole Polytechnique for the warm hospitality in various stages of the collaboration.
The work of C.A. is partially supported by the MIUR-PRIN contract 2017CC72MK-003,
Q.B. is supported by the Deutsche Forschungsgemeinschaft under Germany's Excellence Strategy  EXC 2121 ``Quantum Universe" - 390833306,  C.C. is supported by project `Nucleu' PN-19060101/2019-2022,
whereas E.D. was supported in part by  the ANR grant Black-dS-String ANR-16-CE31-0004-01.

\appendix
\section{Anomaly Polynomials}
\label{appA}
The contribution to gravitational and gauge anomalies from right-handed spin $1/2$ and spin $3/2$ fermions and from anti-selfdual antisymmetric tensors have the expressions (see for ex. \cite{Bilal:2008qx})
\begin{align}
\hat I^{1/2}_{2r+2} &= 2 \pi \left[\hat A(M_{2r})\  \text{ch} (-F) \right]_{2r+2} \ , \\
\hat I^{3/2}_{2r+2} & = 2 \pi \left[\hat A(M_{2r}) \left(\text{tr}\  e^{2i R} -1\right)\ \text{ch}(-F) \right]_{2r+2}\ ,\\
\hat I^A_{2r+2} & = 2 \pi \left[\left(-\frac{1}{2} \right) \frac14\ L(M_{2r}) \right]_{2r+2}\ ,
\end{align}
where one picks the $2r+2$ form after expanding the polynomials. The genus $\hat A$ and the Hirzebruch polynomial $L$ are given by
\be
\begin{split}
\hat A (M_{2r}) &= 1 + \frac{1}{12} \text{tr} R^2 + \frac{1}{360} \text{tr} R^4 + \frac{1}{288} \left(\text{tr} R^2\right)^2\\
&+\frac{1}{5670} \text{tr} R^6 + \frac{1}{4320} \text{tr} R^4\ \text{tr} R^2 + \frac{1}{10368} \left(\text{tr} R^2\right)^3 + \dots
\end{split} \ , \label{aroof}
\ee
\be
\begin{split}
L(M_{2r}) &= 1 -4\times  \frac{1}{6} \text{tr} R^2 +16\times \left[ -\frac{7}{180} \text{tr} R^4 + \frac{1}{72} \left(\text{tr} R^2\right)^2 \right]\\
&+64\times \left[-\frac{31}{2835} \text{tr} R^6 + \frac{7}{1080} \text{tr} R^4\ \text{tr} R^2 - \frac{1}{1296} \left(\text{tr} R^2\right)^3\right] + \dots
\end{split} \ ,
\ee
with $M_{2r}$ being the spacetime manifold. It is easy to see that we can also write
\begin{align}
\hat A(M_{2r})  \left(\text{tr}\  e^{2i R} -1\right) &= 2r-1 + \frac{2r-25}{12} \text{tr} R^2 + \frac{2r+239}{360} \text{tr} R^4 + \frac{2r-49}{288} \left(\text{tr} R^2\right)^2\\
&+ \frac{2r-505}{5670} \text{tr} R^6 + \frac{2r+215}{4320} \text{tr} R^4\ \text{tr} R^2 + \frac{2r-73}{10368} \left(\text{tr} R^2 \right)^3+ \dots
\end{align}
and for the Chern character
\begin{align}
\text{ch} (-F) = n_\psi - \frac12 \text{tr}_\psi F^2 + \frac{1}{4!} \text{tr}_{\psi} F^4 - \frac{1}{6!} \text{tr}_{\psi} F^6 + \dots \ ,
\end{align}
where our conventions for the normalization of $F$ and $R$ are such that we have
\begin{align}
R &= \frac{R_{\text{ref. \cite{Bilal:2008qx}}}}{4 \pi}\ , & F =  \frac{F_{\text{ref. \cite{Bilal:2008qx}}}}{2 \pi} \ .
\end{align}
From the equations above one derives the following anomaly polynomials ($I\equiv \hat I/2\pi$) relevant for the six dimensional orientifold models that we consider
\begin{align}
I_8^{1/2} &= n_\psi \left[\frac{1}{360} \text{tr} R^4 + \frac{1}{288} \left(\text{tr} R^2 \right)^2 \right]  - \frac{1}{24} \text{tr} R^2 \, \text{Tr}_\psi F^2 + \frac{1}{24} \text{Tr}_\psi F^4\ ,\\
I_8^{3/2} & = \frac{245}{360} \text{tr} R^4 - \frac{43}{288} \left(\text{tr} R^2 \right)^2\ ,\\
I_8^A &=   \frac{28}{360} \text{tr} R^4 - \frac{8}{288} \left(\text{tr} R^2 \right)^2\ ,
\end{align}
where we have assumed the gravitino to be uncharged with respect to the gauge group. For the left-handed chirality or selfdual tensor the overall sign of the anomaly polynomial gets flipped. For D1 branes the spectrum contains only spin $1/2$ (symplectic) Majorana-Weyl fermions. In this case the anomaly polynomial has the form
\be
I_4^{1/2} = \frac12 \left(\frac{n_\psi}{12} \text{tr} R^2 - \frac12 \text{Tr}_\psi F^2 \right) \ .
\ee
A more refined version of the anomaly polynomial $I_4$ can be obtained by taking into consideration the contributions from the decomposition into tangent and normal bundle to the worldvolume of the D1 branes. In this case one has \cite{Dudas:2001wd}
\be
I_4 = \hat A(R) \hat A(N)^{-1} \times \left\{
\begin{split}
&\text{ch}_\pm(N) \, \text{Tr}_\psi e^{iG}\\
&\text{tr}\,  e^{iF}\, \text{tr}\,  e^{iG}
\end{split}\ , \right.
\ee
where the first line counts contributions from states charges only under the D1 brane gauge group $G$ and the second line counts contributions from bifundamental representatations. The genus $\hat A$ is given in eq. \eqref{aroof}. For the $I_4$ polynomial one has the following relevant terms in the expansion of the Dirac genus and Chern characters
\begin{align}
\hat A(R) & = 1+\frac1{12} \text{tr} R^2 + \hdots \ ,\\
\hat A(N)^{-1} & = 1- \frac1{12} \text{tr} N^2 + \hdots \ ,\\
\text{ch}_\pm (N) & = 2-\frac12 \text{tr} N^2 \pm \frac12 \chi(N)+\hdots \ ,
\end{align} 
where $\chi(N)$ is the Euler class of the normal bundle. Notice that for Majorana-Weyl fermions, as is the case for D1 branes, one needs to include a factor of $1/2$.

For perturbative D-brane models one can have only antisymmetric, symmetric, adjoint and bifundamental representations of the gauge group, generically being a product of orthogonal, symplectic or unitary factors. The traces of these representations can be expressed in terms of the traces over the fundamental as follows
\begin{table}[h!]
\centering
\begin{tabular}{cc}
{\bf Representation} & {\bf Decomposition}\\
\hline\hline\\[-10pt]
Antisymmetric & $\text{Tr} F^2 = (N-2) \text{tr} F^2$\\[2pt]
$SU/SO/USp(N)$ &  $\text{Tr} F^4  = (N-8) \text{tr} F^4 + 3 \left(\text{tr} F^2 \right)^2$\\[2pt]
\hline\\[-10pt]
Symmetric & $\text{Tr} F^2 = (N+2) \text{tr} F^2$\\[2pt]
$SU/SO/USp(N)$ &  $\text{Tr} F^4  = (N+8) \text{tr} F^4 + 3 \left(\text{tr} F^2 \right)^2$ \\[2pt]
\hline\\[-10pt]
Adjoint & $\text{Tr} F^2 = 2N \text{tr} F^2$ \\[2pt]
$SU(N)$ & $ \text{Tr} F^4  = 2N \text{tr} F^4 + 6 \left(\text{tr} F^2 \right)^2 $\\[2pt]
\hline\\[-10pt]
Bifundamental & $\text{Tr}_{(m,n)} F^2 = m\, \text{tr}_n F^2 + n\, \text{tr}_m F^2$\\[2pt]
$G_1 \times G_2$ & $ \text{Tr}_{(m,n)} F^4  = m\, \text{tr}_n F^4 + n\, \text{tr}_m F^4 + 6\, \text{tr}_m F^2\,  \text{tr}_{n} F^2$\\[2pt]\hline
\end{tabular}
\end{table}
\\
\section{D1 brane Amplitudes for the Supersymmetric $T^4/\mathbb{Z}_2$ Orientifold}
\label{appB}
All our notations and conventions for the open string cylinder and Mobius amplitudes are explained in \cite{Angelantonj:2002ct}, to which we refer the reader for more details.

\subsection{D1-branes at an orbifold fixed point}

We reproduce here the open string vacuum amplitudes corresponding to D1 branes on the $T^4/\mathbb{Z}_2$ orientifold with O9$_-$ and O5$_-$ planes.
\begin{align}
&\mathcal{A}_{11} = \frac14 (r+ \bar r)^2 \left[O_0 (O_4 V_4+V_4 O_4  ) + V_0( O_4 O_4 + V_4  V_4) - S_0(S_4  S_4 + C_4  C_4) - C_0(S_4 C_4 + C_4  S_4) \right]\nonumber\\
&- \frac14(r-\bar r)^2 \left[O_0 (- O_4 V_4+V_4 O_4 ) + V_0( O_4 O_4 - V_4  V_4) - S_0(-S_4  S_4 + C_4  C_4) - C_0(S_4 C_4 - C_4  S_4) \right] \ ,
\end{align}
\begin{align}
&\mathcal{M}_1  = -\frac{r+\bar r}{4}\left [-\hat O_0 (\hat O_4  \hat V_4 + \hat V_4 \hat O_4) + \hat V_0 (\hat O_4 \hat O_4 - \hat V_4 \hat V_4) -\hat S_0 (\hat S_4 \hat S_4 + \hat C_4 \hat C_4) + \hat C_0 (\hat S_4 \hat C_4 + \hat C_4 \hat S_4) \right.\nonumber\\
&\left.-\hat O_0 (\hat O_4  \hat V_4 - \hat V_4  \hat O_4) + \hat V_0 (-\hat O_4  \hat O_4 - \hat V_4 \hat V_4) - \hat S_0 (\hat S_4 \hat S_4-\hat C_4  \hat C_4) + \hat C_0 (-\hat S_4  \hat C_4+\hat C_4 \hat S_4) \right]\ , \label{susyM1}
\end{align}
\begin{align}
&\mathcal{A}_{19}  = \frac12(n+ \bar n)(r+ \bar r) \left[O_0 (S_4  C_4+C_4 S_4) + V_0 (S_4  S_4+ C_4  C_4) - S_0 (O_4  O_4+V_4  V_4) - C_0 (O_4 V_4 + V_4  O_4)     \right]\nonumber\\
 & -\frac12(n- \bar n)(r- \bar r) \left[O_0 (S_4  C_4-C_4 S_4) + V_0 (-S_4  S_4+ C_4  C_4) - S_0 (O_4  O_4-V_4  V_4) - C_0 (-O_4 V_4 + V_4  O_4)     \right] \ ,
\end{align}
\begin{align}
&\mathcal{A}_{15} = \frac12(d+ \bar d) (r+ \bar r) \left[O_0 (S_4 V_4 + C_4  O_4) + V_0 (S_4 O_4+C_4  V_4) - S_0 (O_4  S_4 + V_4  C_4) - C_0 (O_4 C_4 + V_4  S_4) \right]\nonumber\\
 &-\frac12(d- \bar d) (r- \bar r) \left[O_0 (-S_4 V_4 + C_4  O_4) + V_0 (S_4 O_4-C_4  V_4) - S_0 (-O_4  S_4 + V_4  C_4) - C_0 (O_4 C_4 - V_4  S_4) \right] \ .
\end{align}
The contributions to the massless spectrum are found to be
\begin{align}
\mathcal{A}_{11}^{(0)} + \mathcal{M}_1^{(0)}& = r \bar r \left(O_0 V_4 O_4 + V_0  O_4  O_4- S_0 C_4 C_4 - C_0 S_4 C_4 \right)\\
&+\frac{r(r+1)+\bar r (\bar r +1)}{2} \left(O_0 O_4 V_4 -C_0 C_4  S_4 \right)+\frac{r(r-1)+\bar r (\bar r -1)}{2} \left(- S_0 S_4 S_4 \right)\ ,\nonumber\\
\mathcal{A}_{19}^{(0)} &= -(n \bar r+\bar n r) S_0 O_4 O_4\ , \\
\mathcal{A}_{15}^{(0)} &= -(d r + \bar d \bar r) S_0 O_4 S_4 + (d \bar r + \bar d r) (O_0  C_4 O_4 - C_0 O_4 C_4) \ .
\end{align}

\subsection{D1 branes in the bulk}

Displacing the D1 branes in the bulk changes its worldvolume gauge group to $SO(r)$. Indeed, this can be inferred from the corresponding vacuum amplitudes
\begin{align}
\mathcal{A}_{11} & = \frac{r^2}{2} \left[O_0 (O_4  V_4+V_4  O_4) + V_0(O_4  O_4 + V_4  V_4) - S_0 (S_4  S_4 + C_4  C_4) - C_0 (S_4  C_4 + C_4 S_4) \right]\nonumber\\
 & \times \left(W+\frac12 W_{2a}+\frac12 W_{-2a} \right) W^{(3)} \ ,
\end{align}
\
\begin{align}
\mathcal{M}_1 &= - \frac{r}2 \left[-\hat O_0 (\hat O_4 \hat V_4 + \hat V_4 \hat O_4) + \hat V_0 ( \hat O_4  \hat O_4 - \hat V_4  \hat V_4) - \hat S_0 (\hat S_4  \hat S_4 + \hat C_4  \hat C_4) + \hat C_0 (\hat S_4  \hat C_4 + \hat C_4  \hat S_4) \right]\nonumber\\
&+\frac{r}2 \left[-\hat O_0 (-\hat O_4 \hat V_4 + \hat V_4  \hat O_4) + \hat V_0 ( \hat O_4  \hat O_4 + \hat V_4  \hat V_4) - \hat S_0 (- \hat S_4  \hat S_4 + \hat C_4  \hat C_4) + \hat C_0 (\hat S_4  \hat C_4 - \hat C_4  \hat S_4) \right]\nonumber\\
&\times \frac12\left(W_{2a} + W_{-2a} \right) W^{(3)} \ .
\end{align}
Furthermore, for the D1-D9 sector one has
\begin{align}
\mathcal{A}_{19} = r (n+\bar n) \left[O_0 (S_4\, C_4 + C_4 \, S_4) + V_0 (S_4 \, S_4 + C_4\, C_4) - S_0 (O_4 \, O_4 +V_4\, V_4) -C_0 (O_4 \, V_4 + V_4\, O_4) \right] \ ,
\end{align}
whereas $\mathcal{A}_{15}$, corresponding to the D1-D5 sector,  has only massive contributions. The contributions to the massless spectrum are then given by
\begin{align}
\mathcal{A}_{11}^{(0)} + \mathcal{M}_1^{(0) } &= \frac{r(r-1)}{2} \left(V_0  O_4 O_4- S_0 S_4 S_4 - S_0 C_4 C_4 \right)\nonumber\\
& +\frac{r(r+1)}{2} \left(O_0 O_4 V_4 + O_0 V_4 O_4 - C_0 S_4 C_4 - C_0 C_4 S_4 \right)\ ,\\
\mathcal{A}_{19}^{(0)} &= -r(n+ \bar n) S_0 O_4 O_4 \ .
\end{align}

\section{D1 brane Amplitudes for the Non-Supersymmetric $T^4/\mathbb{Z}_2$ Orientifold}
\label{appC}
\subsection{D1 branes at an orbifold fixed point}

We reproduce here the open string vacuum amplitudes corresponding to D1 branes on the brane supersymmetry breaking $T^4/\mathbb{Z}_2$ orientifold with O9$_-$ and O5$_+$ planes. 
\begin{align}
&\mathcal{A}_{11} = \frac14 (d_1+d_2)^2 \left[O_0 (O_4 V_4+V_4 O_4  ) + V_0( O_4 O_4 + V_4  V_4) - S_0 (S_4  S_4 + C_4  C_4) - C_0(S_4 C_4 + C_4  S_4) \right]\nonumber\\
&+ \frac14(d_1-d_2)^2 \left[O_0 (- O_4 V_4+V_4 O_4 ) + V_0( O_4 O_4 - V_4  V_4) - S_0(-S_4  S_4 + C_4  C_4) - C_0(S_4 C_4 - C_4  S_4) \right] \ ,
\end{align}
\begin{align}
&\mathcal{M}_1  = - \frac{(d_1+d_2)}{4} \left[-\hat O_0 (\hat O_4  \hat V_4 + \hat V_4 \hat O_4) + \hat V_0 (\hat O_4 \hat O_4 - \hat V_4 \hat V_4) -\hat S_0 (\hat S_4 \hat S_4 + \hat C_4 \hat C_4) + \hat C_0 (\hat S_4 \hat C_4 + \hat C_4 \hat S_4) \right.\nonumber\\
& \left. - \hat O_0 (-\hat O_4  \hat V_4 +\hat V_4 \hat O_4) +\hat V_0 (\hat O_4 \hat O_4 + \hat V_4 \hat V_4) - \hat S_0 (-\hat S_4 \hat S_4 + \hat C_4 \hat C_4) + \hat C_0 (\hat S_4  \hat C_4 - \hat C_4 \hat S_4) \right]\ , \label{nsusyM}
\end{align}
\begin{align}
&\mathcal{A}_{19}  =\frac12(d_1+d_2)(n_1+n_2) \left[O_0 (S_4 C_4 + C_4 S_4) + V_0 (S_4 S_4 + C_4 C_4)- S_0 (O_4 O_4+V_4  V_4) - C_0 (O_4 V_4+ V_4 O_4) \right]\nonumber\\
&+(d_1-d_2)(n_1-n_2) \left[ O_0 (S_4 C_4 - C_4 S_4) + V_0 (-S_4 S_4 + C_4 C_4) - S_0 (O_4 O_4-V_4  V_4) - C_0 (-O_4 V_4+ V_4 O_4)\right] \ ,
\end{align}
\begin{align}
\mathcal{A}_{1\bar 5} = \frac12(d_1+d_2)(m_1+m_2) \left[O_0 (S_4 O_4 + C_4  V_4) +V_0 (S_4 V_4+C_4 O_4 )-S_0(O_4 C_4+V_4 S_4) - C_0 (O_4 S_4 + V_4  C_4) \right]\nonumber\\
+\frac12(d_1-d_2)(m_1-m_2) \left[O_0 (S_4 O_4 - C_4  V_4) +V_0 (- S_4 V_4+C_4 O_4 )-S_0(O_4 C_4-V_4 S_4) - C_0 (-O_4 S_4 + V_4  C_4) \right]\ .
\end{align}
One obtains the following contributions to the massless spectrum
\begin{align}
\mathcal{A}_{11}^{(0)} & = d_1 d_2 \left(O_0  O_4  V_4  - S_0 S_4  S_4 - C_0 C_4  S_4 \right) + \frac{d_1(d_1-1)+d_2(d_2-1)}2 \left( V_0 O_4 O_4 - S_0 C_4  C_4  \right)\ , \nonumber\\
&+ \frac{d_1(d_1+1)+d_2(d_2+1)}2 \left(O_0 V_4  O_4 - C_0 S_4 C_4\right) \\
\mathcal{A}_{19}^{(0)} &= -(d_1n_1+d_2n_2) S_0 O_4 O_4\ , \\
\mathcal{A}_{1\bar 5}^{(0)} & = (d_1 m_1+d_2m_2) \left(O_0 S_4 O_4 - S_0 O_4 C_4 \right) + (d_1m_2+d_2 m_1) \left(-C_0 O_4 S_4 \right)\ .
\end{align}

\subsection{D1 branes in the bulk}

Displacing the D1 branes in the bulk changes its worldvolume gauge group to $SO(d)$ (instead of $SO(d_1) \times SO(d_2)$). Indeed, this can be inferred from the corresponding vacuum aplitudes
\begin{align}
\mathcal{A}_{11} & = \frac{d^2}{2} \left[O_0 (O_4  V_4+V_4  O_4) + V_0(O_4  O_4 + V_4  V_4) - S_0 (S_4  S_4 + C_4  C_4) - C_0 (S_4  C_4 + C_4 S_4) \right]\nonumber\\
 & \times \left(W+\frac12 W_{2a}+\frac12 W_{-2a} \right) W^{(3)}\ ,
\end{align}
\
\begin{align}
&\mathcal{M}_1 = - \frac{d}2 \left[-\hat O_0 (\hat O_4 \hat V_4 + \hat V_4  \hat O_4) + \hat V_0 ( \hat O_4  \hat O_4 - \hat V_4 \hat V_4) - \hat S_0 (\hat S_4  \hat S_4 + \hat C_4  \hat C_4) + \hat C_0 (\hat S_4  \hat C_4 + \hat C_4  \hat S_4) \right]\nonumber\\
&-\frac{d}2 \left[-\hat O_0 (-\hat O_4 \hat V_4 + \hat V_4  \hat O_4) + \hat V_0 ( \hat O_4  \hat O_4 + \hat V_4  \hat V_4) - \hat S_0 (- \hat S_4  \hat S_4 + \hat C_4  \hat C_4) + \hat C_0 (\hat S_4  \hat C_4 - \hat C_4  \hat S_4) \right]\nonumber\\
&\times \frac12\left(W_{2a} + W_{-2a} \right) W^{(3)} \ .
\end{align}
Furthermore, for the D1-D9 sector one has
\begin{align}
\mathcal{A}_{19} = d (n_1+ n_2) \left[O_0 (S_4 C_4 + C_4  S_4) + V_0 (S_4  S_4 + C_4 C_4) - S_0 (O_4  O_4 +V_4 V_4) -C_0 (O_4  V_4 + V_4 O_4) \right]\ ,
\end{align}
whereas $\mathcal{A}_{15}$, corresponding to the D1-D5 sector,  has only massive contributions. The massless spectrum is then given by
\begin{align}
\mathcal{A}_{11}^{(0)} + \mathcal{M}_1^{(0) } &= \frac{d(d-1)}{2} \left(V_0  O_4 O_4- S_0 S_4 S_4 - S_0 C_4 C_4 \right)\nonumber\\
& +\frac{d(d+1)}{2} \left(O_0 O_4 V_4 + O_0 V_4 O_4 - C_0 S_4 C_4 - C_0 C_4 S_4 \right) \ ,\\
\mathcal{A}_{19}^{(0)} &= -d(n_1+ n_2) S_0 O_4 O_4 \ .
\end{align}


\section{Definitions and Notations for Magnetized/Intersecting Branes; D5$'_a$ Brane Spectra}
\label{appD}
The magnetic fields on the D9 and D5$'$ branes in type I are related to the wrapping numbers in the T-dual version of intersecting D7 branes according to
\be
H_i^\alpha = \frac{m_i^\alpha}{n_i^\alpha v_i} \ .
\ee
As mentioned also in the main text, one uses for simplicity the T-dual  D7 brane language of intersecting numbers, although we will still talk about D9 and D5 branes of type I.
The intersection numbers between D7 brane stacks $\alpha$ and $\beta$ ($I_{\alpha \beta}$),  between stack $\alpha$ and image stack $\beta'$ ($I_{\alpha \beta'}$), and the intersection number between 
brane stack $\alpha$ and all orientifold planes ($I_{\alpha O} $ in the SUSY orientifold and  $\tilde I_{\alpha O}$ for the brane supersymmetry breaking orientifold) have the following expressions
\begin{align}
I_{\alpha \beta} &= \prod_{i=1}^2 (m_i^\alpha n_i^\beta - n_i^\alpha m_i^\beta)\ ,  &  I_{\alpha \beta'} & =  \prod_{i=1}^2 (m_i^\alpha n_i^\beta + n_i^\alpha m_i^\beta) \ ,\\
I_{\alpha O} & = 4(m_1^\alpha  m_2^\alpha +n_1^\alpha n_2^\alpha)\ , & \tilde I_{\alpha O} & = 4(m_1^\alpha  m_2^\alpha -n_1^\alpha n_2^\alpha)\ .
\end{align}
Furthermore we have
\be
S_{\alpha \beta} = \text{number of common fixed points that the branes $\alpha$ and $\beta$ intersect.}
\ee
The wrapping numbers/magnetizations for the various branes are given by
\begin{align}
\text{D9}_\alpha &: (m_1^\alpha, n_1^\alpha) \otimes (m_2^\alpha, n_2^\alpha) \ , \nonumber\\
\text{D5}'_a& :  (m_1^a, n_1^a) \otimes (m_2^a, n_2^a) \ , \nonumber\\
\text{D5}',\text{D9} &: (0,1) \otimes (0,1) \ , \\
\text{D1, D5} & : (1,0) \otimes (1,0)\ ,\nonumber\\
\overline{\text{D5}} &: (-1,0) \otimes (1,0) \ . \nonumber
\end{align}
{\flushleft The Chan-Paton parametrizations of the various branes and the corresponding $\mathbb{Z}_2$ action for the supersymmetric and BSB $T^4/\mathbb{Z}_2$ are given below}
\be
\text{SUSY} :
 \left \{ \begin{split}
\text{D9}_\alpha &: p_\alpha + \bar p_\alpha  \rightarrow i(p_\alpha - \bar p_\alpha)\\
\text{D5}'_a& :  r_a + \bar r_a \rightarrow i(r_a- \bar r_a)\\\
\text{D9} &: n+\bar n \rightarrow i(n- \bar n)\\
\text{D1} & : r+\bar r \rightarrow i(r- \bar r)\\
\text{D5} & : d+ \bar d \rightarrow i(d- \bar d)
\end{split} \right.
\quad \quad \quad \text{BSB} :
 \left \{ \begin{split}
\text{D9}_\alpha &: p_\alpha + \bar p_\alpha  \rightarrow \epsilon_\alpha (p_\alpha + \bar p_\alpha)\\
\text{D5}'_a& :  r_a + \bar r_a \rightarrow \epsilon_a (r_a+ \bar r_a)\\\
\text{D9} &: n_1+n_2 \rightarrow n_1-n_2\\
\text{D1} & : d_1+d_2 \rightarrow d_1-d_2\\
\overline{\text{D5}} & : m_1+m_2 \rightarrow m_1-m_2
\end{split} \right.
\ee
where $\epsilon_a, \epsilon_\alpha = \pm 1$.\\
In Tables \ref{D5spectrum1},\ref{D5spectrum2} we reproduce the massless spectra for magnetized D5$'_a$ branes (BPS with respect to the D9 branes) on the supersymmetric and non-supersymmetric $T^4/\mathbb{Z}_2$ orientifolds respectively.
\begin{table}[h!]
\centering
\begin{tabular}{ccc|c}
{\bf Multiplicity} & {\bf Representation} & {\bf $SO(1,1)\times SU(2)_l \times SU(2)_R \times SO(4)$} & {\bf Sector}\\
\hline\hline\\[-10pt]
1 & $r_a \bar r_a$ & $(0,1,1,1)+(\frac12,1,2,2')_L$ & $aa$ \\[2pt]
1 & $r_a \bar r_a$ & $(1,2,2,1)+(\frac12,2,1,2')_R$ & $aa$ \\[2pt]
\hline\\[-10pt]
$\frac14(I_{aa'} + 4+I_{aO})$ & $\frac{r_a(r_a+1)}{2}$ & $(1,1,1,4)+(\frac12,1,2,2)_R$ & $aa'$\\[2pt]
$\frac14(I_{aa'} + 4-I_{aO})$ & $\frac{r_a(r_a-1)}{2}$ & $(1,1,1,4)+(\frac12,1,2,2)_R$ & $aa'$ \\[2pt]
$\frac14(I_{aa'} + 4-I_{aO})$ & $\frac{r_a(r_a+1)}{2}$ & $(\frac12,2,1,2)_L$ & $aa'$ \\[2pt]
$\frac14(I_{aa'} + 4+I_{aO})$ & $\frac{r_a(r_a-1)}{2}$ & $(\frac12,2,1,2)_L $ & $aa'$ \\[2pt]
\hline\\[-10pt]
$\frac12 (I_{a \beta} - S_{a\beta}) $ & $(r_a, \bar p_\beta)$ & $(\frac12,1,1,2)_L$ & $a\beta, a \neq \beta$\\[2pt]
$\frac12 (I_{a \beta'} + S_{a\beta})$ & $(r_a, p_\beta)$ & $(\frac12,1,1,2)_L$ & $a\beta', a \neq \beta$\\[2pt]
\hline\\[-10pt]
$1 $ & $(r_a, \bar p_\alpha)+(\bar r_a,p_\alpha)$  & $(1,1,2,1)+(\frac12,1,1,2')_R$&  $a\alpha, a = \alpha$ \\[2pt]
$\frac12 (I_{aa'} + 4)$ & $(r_a, p_\alpha)$ & $(\frac12,1,1,2)_L$ & $a\alpha' , a = \alpha$\\[2pt]
\hline\\[-10pt]
$\frac12\left(I_{a5}- S_{a5} \right)$ & $( r_a, d)+(\bar r_a, \bar d) $ & $(\frac12,1,1,1)_L$ & $a5$\\[2pt]
$\frac12\left(I_{a5}+ S_{a5} \right)$ & $ ( r_a, \bar d) + (\bar r_a, d)$ & $(\frac12,1,1,1)_L$ & $a5$\\[2pt]
\hline
\end{tabular}
\caption{Massless spectrum of the BPS magnetized D5$'_a$ brane with non-zero $H_1^a=H_2^a$ for the supersymmetric $T^4/\mathbb{Z}_2$ orientifold.} \label{D5spectrum1}
\end{table}

\begin{table}[h!]
\centering
\begin{tabular}{ccc|c}
{\bf Multiplicity} & {\bf Representation} & {\bf $SO(1,1)\times SU(2)_l \times SU(2)_R \times SO(4)$} & {\bf Sector}\\
\hline\hline\\[-10pt]
1 & $r_a \bar r_a$ & $(0,1,1,1)+(\frac12,1,2,2')_L$ & $aa$ \\[2pt]
1 & $r_a \bar r_a$ & $(1,2,2,1)+(\frac12,2,1,2')_R$ & $aa$ \\[2pt]
\hline\\[-10pt]
$\frac14(-I_{aa'} - 4- I_{aO})$ & $\frac{r_a(r_a+1)}{2}$ & $(1,1,1,4)$ & $aa'$\\[2pt]
$\frac14(-I_{aa'} - 4 + I_{aO})$ & $\frac{r_a(r_a-1)}{2}$ & $(1,1,1,4)$ & $aa'$ \\[2pt]
\hline\\[-10pt]
$\frac14(-I_{aa'} + 4-\tilde I_{aO})$ & $\frac{r_a(r_a+1)}{2}$ & $ (\frac12,1,2,2')_L$ & $aa'$ \\[2pt]
$\frac14(-I_{aa'} + 4+ \tilde I_{aO})$ & $\frac{r_a(r_a-1)}{2}$ & $(\frac12,1,2,2')_L $ & $aa'$ \\[2pt]
$\frac14(-I_{aa'} + 4+ \tilde I_{aO})$ & $\frac{r_a(r_a+1)}{2}$ & $ (\frac12,2,1,2')_R$& $aa'$ \\[2pt]
$\frac14(-I_{aa'} + 4- \tilde I_{aO})$ & $\frac{r_a(r_a-1)}{2}$ & $ (\frac12,2,1,2')_R $ & $aa'$\\[2pt]
\hline\\[-10pt]
$\frac12(-I_{a\beta}+\epsilon_a\epsilon_\beta S_{a\beta})$ & $(r_a, \bar p_\beta)$ & $(\frac12,1,1,2')_R$ & $a\beta, a\neq \beta$\\[2pt]
$\frac12(-I_{a\beta'}+\epsilon_a\epsilon_\beta S_{a\beta})$ & $(r_a, p_\beta)$ & $(\frac12,1,1,2')_R$ & $a\beta', a\neq \beta$\\[2pt]
\hline\\[-10pt]
$\frac12(1+\epsilon_a \epsilon_\alpha) $ & $(r_a, \bar p_\alpha)+(\bar r_a,p_\alpha)$  & $(1,1,2,1)+(\frac12,1,1,2')_R$&  $a\alpha, a = \alpha$ \\[2pt]
$\frac12(1-\epsilon_a \epsilon_\alpha) $ & $(r_a, \bar p_\alpha)+(\bar r_a,p_\alpha)$  & $(\frac12,1,1,2)_L$&  $a\alpha, a = \alpha$ \\[2pt]
$\frac12 (I_{aa'} + \epsilon_a \epsilon_\alpha 4)$ & $(r_a, p_\alpha)$ & $(\frac12,1,1,2')_R$ & $a\alpha' , a = \alpha$\\[2pt]
\hline\\[-10pt]
$\frac12(-I_{a9}+\epsilon_aS_{a9})$ & $(r_a, n_1)$ & $(\frac12,1,1,2')_R$ & $a9$\\[2pt]
$\frac12(-I_{a9}-\epsilon_aS_{a9})$ & $(r_a, n_2)$ & $(\frac12,1,1,2')_R$ & $a9$\\[2pt]
\hline\\[-10pt]
$\frac12(I_{a5}+\epsilon_a S_{a5})$ & $(r_a+ \bar r_a, m_1)$ & $(\frac12,1,1,1)_R$ & $a5$\\[2pt]
$\frac12(I_{a5}-\epsilon_a S_{a5})$ & $(r_a+\bar r_a,  m_2)$ & $(\frac12,1,1,1)_R$ & $a5$\\[2pt]
\hline
\end{tabular}
 \caption{Massless spectrum of the BPS magnetized D5$'_a$ brane with non-zero $H_1^a=-H_2^a$ for the brane supersymmetry breaking  $T^4/\mathbb{Z}_2$ orientifold.}\label{D5spectrum2}
\end{table}

\pagebreak

In Tables \ref{D5spectrum3},\ref{D5spectrum4} we reproduce the massless spectra for magnetized D5'$_a$ branes (non-BPS with respect to the D9 branes but stable) on the supersymmetric and non-supersymmetric $T^4/\mathbb{Z}_2$ orientifolds respectively.
\begin{table}[h!]
\centering
\begin{tabular}{ccc|c}
\hline\\[-10pt]
{\bf Multiplicity} & {\bf Representation} & {\bf Characters} & {\bf Sector}\\
\hline\hline\\[-10pt]
1 & $r_a \bar r_a$ & $ (0,1,1,1)+(\frac12,1,2,2')_L$ & $aa$\\[2pt]
1 & $r_a \bar r_a$ & $(1,2,2,1)+(\frac12,2,1,2')_R$ & $aa$\\[2pt]
\hline\\[-10pt]
$\frac14(-I_{aa'} + 4-\tilde I_{aO})$ & $\frac{r_a(r_a+1)}{2}$ & $(1,1,1,4)$ & $aa'$\\[2pt]
$\frac14(-I_{aa'} + 4+ \tilde I_{aO})$ & $\frac{r_a(r_a-1)}{2}$ & $(1,1,1,4) $ & $aa'$ \\[2pt]
\hline\\[-10pt]
$\frac14(-I_{aa'} - 4-I_{aO})$ & $\frac{r_a(r_a+1)}{2}$ & $ (\frac12,1,2,2')_L$ & $aa'$\\[2pt]
$\frac14(-I_{aa'} - 4+I_{aO})$ & $\frac{r_a(r_a-1)}{2}$ & $ (\frac12,1,2,2')_L$ & $aa'$ \\[2pt]
$\frac14(-I_{aa'} - 4+I_{aO})$ & $\frac{r_a(r_a+1)}{2}$ & $ (\frac12,2,1,2')_R$ & $aa'$ \\[2pt]
$\frac14(-I_{aa'} - 4-I_{aO})$ & $\frac{r_a(r_a-1)}{2}$ & $ (\frac12,2,1,2')_R$ & $aa'$ \\[2pt]
\hline\\[-10pt]
$\frac12 (-I_{a \beta} + S_{a\beta}) $ & $(r_a, \bar p_\beta)$ & $(\frac12,1,1,2')_R$ & $a\beta$\\[2pt]
$\frac12 (-I_{a \beta'} - S_{a\beta})$ & $(r_a, p_\beta)$ & $(\frac12,1,1,2')_R$ & $a\beta'$\\[2pt]
\hline\\[-10pt]
$\frac12\left(I_{a5}- S_{a5} \right)$ & $( r_a, d)+(\bar r_a, \bar d) $ & $(\frac12,1,1,1)_L$ & $a5$\\[2pt]
$\frac12\left(I_{a5}+ S_{a5} \right)$ & $ ( r_a , \bar d) + (\bar r_a ,  d)$ & $(\frac12,1,1,1)_L$ & $a5$\\[2pt]
\hline
\end{tabular}
 \caption{Massless spectrum of the non-BPS magnetized D5$'_a$ brane with $H_1^a=-H_2^a \neq0$ for the supersymmetric $T^4/\mathbb{Z}_2$ orientifold.}\label{D5spectrum3}
\end{table}

\begin{table}[h!]
\centering
\begin{tabular}{ccc|c}
\hline\\[-10pt]
{\bf Multiplicity} & {\bf Representation} & {\bf Characters} & {\bf Sector}\\
\hline\hline\\[-10pt]
1 & $r_a \bar r_a$ & $ (0,1,1,1)+(\frac12,1,2,2')_L$ & $aa$\\[2pt]
1 & $r_a \bar r_a$ & $(1,2,2,1)+(\frac12,2,1,2')_R$ & $aa$\\[2pt]
\hline\\[-10pt]
$\frac14(I_{aa'} - 4+ \tilde I_{aO})$ & $\frac{r_a(r_a+1)}{2}$ & $(1,1,1,4)$ & $aa'$\\[2pt]
$\frac14(I_{aa'} - 4 -\tilde I_{aO})$ & $\frac{r_a(r_a-1)}{2}$ & $(1,1,1,4)$ & $aa'$ \\[2pt]
\hline\\[-10pt]
$\frac14(I_{aa'} - 4-\tilde I_{aO})$ & $\frac{r_a(r_a+1)}{2}$ & $ (\frac12,2,1,2)_L$ & $aa'$ \\[2pt]
$\frac14(I_{aa'} - 4+ \tilde I_{aO})$ & $\frac{r_a(r_a-1)}{2}$ & $ (\frac12,2,1,2)_L$ & $aa'$ \\[2pt]
$\frac14(I_{aa'} - 4+ \tilde I_{aO})$ & $\frac{r_a(r_a+1)}{2}$ & $ (\frac12,1,2,2)_R$ & $aa'$ \\[2pt]
$\frac14(I_{aa'} - 4- \tilde I_{aO})$ & $\frac{r_a(r_a-1)}{2}$ & $ (\frac12,1,2,2)_R$ & $aa'$\\[2pt]
\hline\\[-10pt]
$\frac12(I_{a\beta}-\epsilon_a\epsilon_\beta S_{a\beta})$ & $(r_a, \bar p_\beta)$ & $(\frac12,1,1,2)_L$ & $a\beta$\\[2pt]
$\frac12(I_{a\beta'}-\epsilon_a\epsilon_\beta S_{a\beta})$ & $(r_a, p_\beta)$ & $(\frac12,1,1,2)_L$ & $a \beta'$\\[2pt]
\hline\\[-10pt]
$\frac12(I_{a9}-\epsilon_aS_{a9})$ & $(r_a, n_1)$ & $(\frac12,1,1,2)_L$ & $a9$\\[2pt]
$\frac12(I_{a9}+\epsilon_aS_{a9})$ & $(r_a, n_2)$ & $(\frac12,1,1,2)_L$ & $a9$\\[2pt]
\hline\\[-10pt]
$\frac12(I_{a5}+\epsilon_a S_{a5})$ & $(r_a+ \bar r_a, m_1)$ & $(\frac12,1,1,1)_R$ & $a5$\\[2pt]
$\frac12(I_{a5}-\epsilon_a S_{a5})$ & $(r_a+\bar r_a,  m_2)$ & $(\frac12,1,1,1)_R$ & $a5$\\[2pt]
\hline
\end{tabular}
 \caption{Massless spectrum of the non-BPS magnetized D5$'_a$ brane with $H_1^a=H_2^a \neq 0$ for the brane supersymmetry breaking  $T^4/\mathbb{Z}_2$ orientifold.}\label{D5spectrum4}
\end{table}

\pagebreak


\section{Extra Models}
\label{appE}


\begin{flushleft}
{\bf BSB in the bulk $SO(16) \times USp(16)$}
\end{flushleft}

We consider a model with continuous Wilson lines on the D9 branes and $\overline{\text{D5}}$ branes displaced in the bulk (T-dual to the intersecting (orthogonal) D7 branes model presented in \cite{Angelantonj:2011hs}) such that the gauge group of the BSB orientifold becomes $SO(16) \times USp(16)$.
The open string spectrum is summarized in Table \ref{bsb-4}.
\begin{table}[h!]
\centering
\begin{tabular}{ccc}
{\bf Field/Multiplet} &  {\bf Multiplicity} & {\bf Representation}
  \\\hline\hline\\[-10pt]
  Gauge Multiplet (L) & $1$ & (120,1)  
  \\[2pt]
Vector Boson &  $1$ & (1,136)  
  \\[2pt] 
  Weyl Fermion (L) & $1$ & (1,120)  
  \\[2pt] \hline\\[-10pt]
 Hypermultiplet (R) &  $1$ & (136,1)  
  \\[2pt]
Scalar &  $4$ & (1,120)  
  \\[2pt]
Weyl Fermion (R) &  $1$ & (1,136)   
  \\[2pt]
Hypermultiplet (L) &  $1$ & (16,16) 
\\ \hline
\end{tabular}
\caption{Open string spectrum for the $SO(16) \times USp(16)$ model.} \label{bsb-4}
\end{table}

From the spectrum, the anomaly polynomial in factorized form is found to be
\begin{align}
I_8 &= \frac{1}{16} \left(\text{tr} F_1^2 - \text{tr} F_2^2 \right)^2 - \frac{1}{16} \left(- 4 \text{tr} R^2 + \text{tr} F_1^2 + \text{tr} F_2^2 \right)^2 \ .
\end{align}
It is easy to show from above that one obtains the same constraints for $J$ as for the model at fixed points given in eq. \eqref{JBSB} and hence it is not possible to define $J$.

\begin{flushleft}
{\bf BSB with Wilson Lines $SO(8)^4 \times USp(8)^4$}
\end{flushleft}
We now consider a BSB model with discrete Wilson lines on the D9 and $\overline{\text{D5}}$'s distributed on fixed points such that the gauge group becomes $SO(8)^4_9 \times USp(8)_{\bar 5}^4$.
The open string spectrum of this model is given in Table \ref{bsb-5}.
\begin{table}[h!]
\begin{center}
\begin{tabular}{ccc}
{\bf Field/Multiplet} & {\bf Multiplicity} & {\bf Representation}\\
\hline\hline\\[-10pt]
$A_\mu$ &1 & $(\underline{28, 1^3},1^4) + (1^4,\underline{36,1^3})$\\[2pt]
$\chi_L$ &1 & $(\underline{28, 1^7})$\\[2pt]
\hline\\[-10pt]
Hyper Multiplets (R) &1& $(8, 8,1^6) + (1^2,8,8,1^4)$\\[2pt]
\ & 1 &$(1^4,8,8,1^2)+(1^6,8,8)$\\[2pt]
\hline\\[-10pt]
Majorana-Weyl Fermions (L) &1 & $(8,1^3,8,1^3)+(8,1^5,8,1)$\\[2pt]
\ & 1 & $(1,8,1^3,8,1^2)+(1,8,1^5,8)$\\[2pt]
\ & 1 & $(1^2,8,1,8,1^3)+(1^2,8,1^3,8,1)$\\[2pt]
\ & 1 & $(1^3,8,1,8,1^2)+(1^3,8,1^3,8)$\\[2pt]
\hline\\[-10pt]
Scalars &2 & $\hdots$\\[2pt]
\hline
\end{tabular}
\end{center}
\caption{Open string spectrum for the $SO(8)^4 \times USp(8)^4$ model.} \label{bsb-5}
\end{table}

From the spectrum, the anomaly polynomial in factorized form is found to be
\begin{align}
I_8 =&\frac{1}{64} \left(\text{tr} F_1^2+ \text{tr} F_2^2+ \text{tr} F_3^2 +\text{tr} F_4^2 - \text{tr} F_5^2 - \text{tr} F_6^2- \text{tr} F_7^2 - \text{tr} F_8^2 \right)^2\nonumber\\
&-\frac{1}{64} \left(-8\, \text{tr} R^2+\text{tr} F_1^2+ \text{tr} F_2^2+ \text{tr} F_3^2 +\text{tr} F_4^2 + \text{tr} F_5^2 + \text{tr} F_6^2+ \text{tr} F_7^2+ \text{tr} F_8^2 \right)^2 \nonumber\\
&-\frac{1}{128} \left( \text{tr} F_1^2- \text{tr} F_2^2+ \text{tr} F_3^2 -\text{tr} F_4^2 +4[\text{tr} F_5^2 - \text{tr} F_6^2] \right)^2\nonumber\\
&-\frac{1}{128} \left( \text{tr} F_1^2- \text{tr} F_2^2+ \text{tr} F_3^2 -\text{tr} F_4^2 +4[\text{tr} F_7^2 - \text{tr} F_8^2] \right)^2\nonumber\\
&-\frac{8}{128} \left(\text{tr} F_1^2- \text{tr} F_2^2- \text{tr} F_3^2 +\text{tr} F_4^2 \right)^2 -\frac{6}{128} \left(\text{tr} F_1^2- \text{tr} F_2^2+ \text{tr} F_3^2 -\text{tr} F_4^2 \right)^2 \ .
\end{align}
Again, one cannot define $J$ for this model. Indeed, it is easy to see from the factorization above that one arrives at the same constraints in eq. \eqref{JBSB} which do not have a solution.

\end{document}